# WEAK VISCOELASTIC NEMATODYNAMICS OF MAXWELL TYPE

**Arkady I. Leonov[a1] , Valery S. Volkov[b]**

[a)] *Department of Polymer Engineering, The University of Akron, Akron, Ohio 44325-0301, USA.*

[b)] *Laboratory of Rheology, Institute of Petrochemical Synthesis, Russian Academy of Sciences, Leninsky Pr., 29, Moscow 117912 Russia.*

**Abstract**

A constitutive theory for weak viscoelastic nematodynamics of Maxwell type is developed using the standard local approach of non-equilibrium thermodynamics. Along with particular viscoelastic and nematic kinematics, the theory uses the weakly elastic potential proposed by de Gennes for nematic solids and the LEP constitutive equations for viscous nematic liquids, while ignoring the Frank (orientation) elasticity and inertia effects. In spite of many basic parameters, algebraic properties of nematic operations investigated in Appendix, allowed us to reveal a general group structure of the theory and present it in a simple form. It is shown that the evolution equation for director is also viscoelastic. An example of magnetization clarifies the situation with non-symmetric stresses. When the sources of stress asymmetry are absent, the theory is simplified and its relaxation properties are described by a symmetric subgroup of nematic algebraic operations. A purely linear constitutive behavior exemplifies the symmetric situation.

[1] Corresponding author.
*E-mail address*: leonov@uakron.edu (A.I. Leonov)

## 1. Introduction

For various liquid crystalline (LC) materials, the internal rotational degree of freedom results in occurrence of internal couples and non-symmetry of stresses. Additionally, for the LC polymers (LCP's) a partial flexibility of polymer chains has also to be taken into account. It means that along with nematic phenomena, typical for low molecular weight LC's, the viscoelasticity of LCP's could also play an important role. In general, the viscoelastic characteristics of these systems, being dependent on the direction of measurement relative to the direction of relaxation anisotropy, are defined by essentially larger number of constitutive scalar parameters, as compared to their isotropic polymeric counterparts. Their experimental determination presents a challenging problem.

Two types of theories, continuum and molecular, attacked the problem of modeling LCP properties. While the continuum theories try to establish a general framework with minimum assumptions about the molecular structures of LCP's, the molecular approaches employ very specific assumptions of structure of these polymers, but unlike the continuum approaches, operate with few molecular parameters. These theories, however well separated, are not contradictory but supplemental.

Volkov and Kulichikhin (1990, 2000a,b) proposed a continuum approach to weak anisotropic viscoelasticity of Maxwell type with internal rotations, based on symmetry arguments. Pleiner and Brand (1991, 1992) developed a thermodynamic theory for linear anisotropic viscoelasticity for LC polymers. Rey (1995a,b) also applied a thermodynamic approach for describing weak nonlinear phenomena in flow of LCP's, However, he obtained doubtful results regarding occurrence of asymmetric stress in his scheme. Terentjev and Warner (2001) developed a thermodynamic theory of solid viscoelasticity for LC elastomers, based on the Kelvin-Voigt type of nematic modeling (see also Fradkin *et al* (2003) and the recent text by Warner and Terentjev (2003)). Leonov and Volkov (2002,2003a,b) initiated non-equilibrium thermodynamic studies of nonlinear nematic viscoelasticity for various polymer systems of different rigidity, such as LCP, LC elastomers and precursors of polymer nanocomposites.

A lot of theoretical effort was made to develop molecular theories that could model the lyotropic LC polymers.. The theories by Marrucci and Greco (1993), Larson and co-authors (1998), and Feng et al (2000) typically use and elaborate further the molecular long rigid rod approach, proposed by Doi (1981) and developed more in the text by Doi and Edwards (1986). Also, B. Edwards *et al* (1990) applied the Poisson-Bracket continuum approach for developing constitutive equations (CE's) for LCP's [see also the text by Beris and Edwards (1999)]. This theory is reduced to the Doi theory in the homogeneous (monodomain) limit. Larson and Mead (1989) extended the Ericksen flow theory of LMV LC's to viscoelastic case, using linear viscoelastic operators instead of Ericksen's viscosities. All the abovementioned theories in this paragraph employ the same state variables as in case of LMW liquid crystals, i.e. the director $\underline{n}$ (or the second rank order tensor), and the director's space gradient $\underline{\nabla n}$. Additionally, we should mention the Rouse-like molecular theories of LCP's by Volkov and Kulichikhin (1994) and Long and Morse (2002) that take into account the partial flexibility of LCP macromolecular chains and [in paper by Volkov and Kulichikhin (1994)] anisotropic macromolecular environment. These papers used the approach by Larson and Mead (1989) for describing the linear viscoelastic phenomena in LCP's.

The present work develops a thermodynamic theory of non-symmetric weak viscoelastic nematodynamics of Maxwell type. The assumption of small transient (elastic) strains and relative rotations, employed in the theory, seems to be valid for LCP's with rigid enough macromolecules. The theory nontrivially combines CE's for weak nematic solids resulted from elastic potential proposed by de Gennes (1980), and the LEP CE's for viscous nematodynamics [e.g. see de Gennes and Prost (1974)]. This theory utilizes a specific viscoelastic and nematic kinematics and ignores the Frank (orientational) elasticity and inertia effects. In spite of many basic parameters, a group of linear nematic operations, introduced and analyzed in Appendix, displays the general structure of the theory in a simple way. When all the sources of stress asymmetry such as rotational inertia, external fields, Frank (orientational) elasticity, etc, are negligible, the theory is simplified and its relaxation properties are described by a subgroup of symmetric nematic operations. In this case, relaxation equations and orientational equation for director are similar to those obtained earlier by Leonov and Volkov (2002, 2003a,b).

It should also be mentioned, that in theories describing possible soft nematic deformation modes the number of material parameters is highly reduced. These soft nematic modes were completely analyzed for elastic and viscous nematics by Leonov and Volkov (2004a,b). Such a theory has yet to be developed for viscoelastic case.

## 2. Non-symmetric tensor formulation

Bearing in mind that the main objective of this paper is elaborating viscoelastic, nematic CE's we avoid discussing here the well-known general relations for balance of moment of momentum and related rotational inertia effects. An interested reader could find the references in above cited monographs and in more detail, in work by Leonov and Volkov (2002).

### *2.1. Kinematical relations*

We use in the following the viscoelastic kinematics based on the decomposition, $\underline{\underline{F}} = \underline{\underline{F}}_e \cdot \underline{\underline{F}}_p$, of full strain gradient $\underline{\underline{F}}$ into elastic (transient) $\underline{\underline{F}}_e$ and inelastic (viscous) $\underline{\underline{F}}_p$ parts [e.g. see Leonov (1987)]. Differentiating this equation with respect to time with further right multiplication by $\underline{\underline{F}}$, yields the Eulerian kinematical rate equation: $\underline{\underline{\nabla \mathrm{v}}} \equiv \dot{\underline{\underline{F}}} \cdot \underline{\underline{F}}^{-1} = \dot{\underline{\underline{F}}}_e \cdot \underline{\underline{F}}_e^{-1} + \underline{\underline{F}}_e \cdot \dot{\underline{\underline{F}}}_p \cdot \underline{\underline{F}}_p^{-1} \cdot \underline{\underline{F}}_e^{-1}$. Here $\underline{\underline{\nabla \mathrm{v}}}$ is the velocity gradient tensor. We now assume the elastic transient strain $\underline{\underline{\varepsilon}}$ and (body) elastic rotation $\underline{\underline{\Omega}}_e^b$ to be small, i.e. $\underline{\underline{F}}_e \approx \underline{\underline{\delta}} + \underline{\underline{\varepsilon}} + \underline{\underline{\Omega}}_e^b$, where $\underline{\underline{\delta}}$ is the unit tensor. Inserting these formulae into the above Eulerian kinematical relation results in the simplified rate equations:

$$\overset{0}{\underline{\underline{\varepsilon}}} + \underline{\underline{e}}_p = \underline{\underline{e}} \qquad (1)$$

$$\overset{0}{\underline{\underline{\Omega}}}{}_e^b + \underline{\underline{\omega}}_p^b = \underline{\underline{\omega}}^b \qquad (2_1)$$

Here the symbol $^0$ stands for Jaumann co-rotation derivative, $\underline{\underline{e}} = 1/2[\underline{\underline{\nabla \mathrm{v}}} + (\underline{\underline{\nabla \mathrm{v}}})^T]$ and $\underline{\underline{\omega}}^b = 1/2[\underline{\underline{\nabla \mathrm{v}}} - (\underline{\underline{\nabla \mathrm{v}}})^T]$ are respectfully the full strain rate and vorticity of "body", so that $\underline{\underline{\nabla \mathrm{v}}} = \underline{\underline{e}} + \underline{\underline{\omega}}^b$, whereas $\underline{\underline{e}}_p$ and $\underline{\underline{\omega}}_p^b$ being respectfully the irreversible (viscous) strain rate and vorticity. Kinematical rate equations (1) and ($2_1$) could approximately describe only small elastic strain/vorticity superimposed on large inelastic strain/vorticity [e.g. see Gorodtsov and Leonov (1968)] This however, is a quite realistic feature for flows of weakly elastic nematic liquids. For simplicity we will consider in the following the incompressible situation when the tensors $\underline{\underline{\varepsilon}}$, $\underline{\underline{e}}_p$ and $\underline{\underline{e}}$ are traceless.

We now introduce in kinematical relations (1), ($2_1$) an additional rate equation describing internal rotations for nematic continua:

$$\overset{0}{\underline{\underline{\Omega}}}{}_e^i + \underline{\underline{\omega}}_p^i = \underline{\underline{\omega}}^i . \qquad (2_2)$$

Here, following Pleiner and Brand (1991, 1992) we decomposed the total tensor of internal vorticity $\underline{\underline{\omega}}^i$ into the rate of reversible rotation $\overset{0}{\underline{\underline{\Omega}}}{}_e^i$ and irreversible vorticity $\underline{\underline{\omega}}_p^i$. Extracting ($2_2$) from ($2_1$) results in the final equation describing the *relative rotations* in weakly nonlinear nematic viscoelastic liquid:

$$\overset{0}{\underline{\underline{\Omega}}}{}_e + \underline{\underline{\omega}}_p = \underline{\underline{\omega}} \qquad \left(\underline{\underline{\Omega}}_e = \underline{\underline{\Omega}}_e^b - \underline{\underline{\Omega}}_e^i,\ \underline{\underline{\omega}}_p = \underline{\underline{\omega}}_p^b - \underline{\underline{\omega}}_p^i,\ \underline{\underline{\omega}} = \underline{\underline{\omega}}^b - \underline{\underline{\omega}}^i\right). \qquad (2_3)$$

Additionally, there is the well-known kinematical relation, describing the rigid rotations of director $\underline{n}$ as:

$$\underline{\underline{\omega}} \cdot \underline{n} = \overset{0}{\underline{\underline{\Omega}}}{}_e \cdot \underline{n} + \underline{\underline{\omega}}_p \cdot \underline{n} = \overset{0}{\underline{n}} . \qquad (3)$$

We finally introduce the kinematical variables, convenient for characterizing a combined effect of viscoelastic deformations and relative rotations:

$$\overset{0}{\underline{\underline{\Gamma}}}{}_e + \underline{\underline{\gamma}}_p = \underline{\underline{\gamma}} \qquad \left(\underline{\underline{\Gamma}}_e = \underline{\underline{\varepsilon}} + \underline{\underline{\Omega}}_e,\ \underline{\underline{\gamma}}_p = \underline{\underline{e}}_p + \underline{\underline{\omega}}_p,\ \underline{\underline{\gamma}} = \underline{\underline{e}} + \underline{\underline{\omega}}\right). \qquad (4)$$

Note that kinematical tensors in (4) written as sums of symmetric and anti-symmetric components, should not be viewed as deformation gradients and velocity gradients, because their asymmetric parts describe the *relative* rotations.

### 2.2. Thermodynamics and general constitutive relations

To describe the *quasi-equilibrium effects* in weak nematic viscoelasticity we will use the de Gennes type of potential (Helmholtz free energy density):

$$\begin{aligned} f &= 1/2G_0\left|\underline{\underline{\varepsilon}}\right|^2 + G_1\underline{nn}:\underline{\underline{\varepsilon}}^2 + G_2(\underline{nn}:\underline{\underline{\varepsilon}})^2 - 2G_3\underline{nn}:(\underline{\underline{\varepsilon}}\cdot\underline{\underline{\Omega}}_e) - G_5\underline{nn}:\underline{\underline{\Omega}}_e^2 + f_m \\ f_m &= -1/2\underline{\underline{\chi}}(\underline{n}):\underline{H}\underline{H}, \quad \underline{\underline{\chi}}(\underline{n}) = \chi_\perp\underline{\underline{\delta}} + \chi_a\underline{nn} \end{aligned} \tag{5}$$

Here $G_k$ are the elastic moduli, $\underline{H}$ is the magnetic field, $f_m$ is the "magnetic" contribution in the free energy, $\underline{\underline{\chi}}(\underline{n})$ is the susceptibility tensor, $\chi_\parallel$ and $\chi_\parallel$ are the susceptibilities parallel and perpendicular to the director, with $\chi_a = \chi_\parallel - \chi_\perp$ being the magnetic anisotropy. In case of diamagnetism, $\chi_\parallel$ and $\chi_\parallel$ are negative. Because of a typical weak magnetization of the considered diamagnetic liquid, the effect of magnetic field on constitutive parameters $G_k$ in (5) can be neglected.

The magnetic body force ("effective magnetic field") is defined as:

$$\underline{h} = -\partial f/\partial\underline{n} \approx -\partial f_m/\partial\underline{n} = \chi_a\underline{H}(\underline{n}\cdot\underline{H}). \tag{$6_1$}$$

In ($6_1$) we employed the smallness of elastic (transient) strains and rotations, and assumed that the corresponding terms are considerably less than these caused by the magnetic field. Along with vector $\underline{h}$ it is also convenient to introduce the asymmetric effective tensor magnetic field $\underline{\underline{h}}_n$ defined as:

$$\underline{\underline{h}}_n \equiv 1/2(\underline{n}\cdot\underline{h} - \underline{h}\cdot\underline{n}). \tag{$6_2$}$$

We now use the well-known formula for the *entropy production* $P_s$ in the system:

$$TP_s = -\underline{q}\cdot\underline{\nabla}T + \underline{\underline{\sigma}}^s:\underline{\underline{e}} + \underline{\underline{\sigma}}^a:\underline{\underline{\omega}} - df\big|_T/dt. \tag{$7_1$}$$

Here $\underline{q}$ is the thermal flux, $T$ is the temperature, $\underline{\underline{\sigma}}^s$ and $\underline{\underline{\sigma}}^a$ are the symmetric and asymmetric parts of the extra stress tensor, respectively. Due to the second law $P_s \geq 0$, i.e. $P_s$ is strictly positive for non-equilibrium processes and vanishes in the equilibrium. Calculating the last term in ($7_1$) with the use of equations (1), ($2_3$), (3)-(6) yields:

$$\begin{aligned} df/dt\big|_T &= \partial f/\partial\underline{\underline{\varepsilon}}:(\underline{\underline{e}} - \underline{\underline{e}}_p) + \partial f/\partial\underline{\underline{\Omega}}_e:(\underline{\underline{\omega}} - \underline{\underline{\omega}}_p) + \partial f/\partial\underline{n}\cdot\overset{0}{\underline{n}} \\ \partial f/\partial\underline{n}\cdot\overset{0}{\underline{n}} &= \underline{h}\cdot\underline{\underline{\omega}}\cdot\underline{n} = (\underline{nh}):\underline{\underline{\omega}} = \underline{\underline{h}}_n:\underline{\underline{\omega}} \end{aligned} \tag{$7_2$}$$

Substituting ($7_2$) into ($7_1$) results in:

$$TP_s = -\underline{q}\cdot\underline{\nabla}T + \underline{\underline{\sigma}}_p^s : \underline{\underline{e}} + \underline{\underline{\sigma}}_p^a : \underline{\underline{\omega}} + \underline{\underline{\sigma}}_e^s : \underline{\underline{e}}_p + \underline{\underline{\sigma}}_e^a : \underline{\underline{\omega}}_p ,$$

$$\underline{\underline{\sigma}}_p^s \equiv \underline{\underline{\sigma}}^s - \underline{\underline{\sigma}}_e^s,\ \ \underline{\underline{\sigma}}_p^a \equiv \underline{\underline{\sigma}}^a - \underline{\underline{h}}_n - \underline{\underline{\sigma}}_e^a,\ \ \underline{\underline{\sigma}}_e^s = \partial f / \partial \underline{\underline{\varepsilon}},\ \ \underline{\underline{\sigma}}_e^a = \partial f / \partial \underline{\underline{\Omega}}_e . \qquad (7_3)$$

Here $\underline{\underline{\sigma}}_e^s$ and $\underline{\underline{\sigma}}_e^a$ are equilibrium symmetric and asymmetric (in the absence of magnetic field) parts of the extra stress tensor. Note that an additional component of asymmetric stress, $\underline{\underline{h}}_n$, commonly ignored in existing theories, occurs automatically as soon as the "magnetic" contribution $f_m$ is included in the free energy.

Expression ($7_3$) for the entropy production demonstrates the various possible sources of dissipation in nematic viscoelastic liquid: (i) non-isothermality, (ii) the power produced by the non-equilibrium extra stress, $\underline{\underline{\sigma}}_p^s + \underline{\underline{\sigma}}_p^a$, on the total strain rate $\underline{\underline{e}}$ and relative vorticity $\underline{\underline{\omega}}$, and (iii) the power produced by the equilibrium extra stress, $\underline{\underline{\sigma}}_e^s + \underline{\underline{\sigma}}_e^a$, on the irreversible strain rate $\underline{\underline{e}}_p$ and relative vorticity $\underline{\underline{\omega}}_p$.

The general quasi-linear constitutive relation between the thermodynamic forces $\{\underline{\nabla}T, \underline{\underline{\sigma}}_p^s, \underline{\underline{\sigma}}_p^a, \underline{\underline{\sigma}}_e^s, \underline{\underline{\sigma}}_e^a\}$ and fluxes $\{\underline{q}, \underline{\underline{e}}, \underline{\underline{\omega}}, \underline{\underline{e}}_p, \underline{\underline{\omega}}_p\}$, with account for Onsager symmetry, can schematically be represented via the "grand matrix" $\{\mathbf{A}^{mn}\}$ as:

$$\underline{q} = -\underline{\underline{k}}\cdot\underline{\nabla}T, \qquad \begin{pmatrix} \underline{\underline{\sigma}}_p^s \\ \underline{\underline{\sigma}}_p^a \\ \underline{\underline{\sigma}}_e^s \\ \underline{\underline{\sigma}}_e^a \end{pmatrix} = \begin{pmatrix} \mathbf{A}^{11} & \mathbf{A}^{12} & \mathbf{A}^{13} & \mathbf{A}^{14} \\ \mathbf{A}^{12\mathbf{T}} & \mathbf{A}^{22} & \mathbf{A}^{23} & \mathbf{A}^{24} \\ \mathbf{A}^{13} & \mathbf{A}^{23\mathbf{T}} & \mathbf{A}^{33} & \mathbf{A}^{34} \\ \mathbf{A}^{14\mathbf{T}} & \mathbf{A}^{24} & \mathbf{A}^{34\mathbf{T}} & \mathbf{A}^{44} \end{pmatrix} \bullet \begin{pmatrix} \underline{\underline{e}} \\ \underline{\underline{\omega}} \\ \underline{\underline{e}}_p \\ \underline{\underline{\omega}}_p \end{pmatrix}. \qquad (*)$$

Here $\underline{\underline{k}}(\underline{n}, T)$ is the positively definite, symmetric, second rank heat conductivity tensor, and $\mathbf{A}_{ijkl}^{mn}(T, \underline{n})$ are fourth rank traceless kinetic tensors of phenomenological coefficients, having dimensionality of viscosity, whose upper and lower indexes characterize their matrix and tensor properties, respectively. The additional upper symbol $^{\mathbf{T}}$ means transposition of the first and second pairs of tensor indexes.

General properties of symmetry of the kinetic tensors, similar to those revealed in Appendix, are summarized as follows. When the first (or second) matrix index is odd, the kinetic tensor is symmetric in transposition of the first (or second) pair of tensor indexes. When the first (or second) matrix index is even, the kinetic tensor is skew symmetric in transposition of the first (or second) pair of tensor indexes. When the sum of the first and second matrix indexes is even, the kinetic tensor is symmetric in transposition of the first and second pairs of indexes. Demanding the grand matrix $\{\mathbf{A}^{mn}\}$ in the right-hand side of (*) to be positively definite, guaranties that the entropy production is positive, allowing also the inverse constitutive relations to exist.

Although the general CE (*) for weakly viscoelastic nematics seems to be intractable, it still contains two relatively simple particular cases:

(i) The CE for liquid viscoelastic nematics of Maxwell type, when in ($7_3$) $\underline{\underline{\sigma}}_p^s = 0, \quad \underline{\underline{\sigma}}_p^a = 0$;

(ii) The CE for solid viscoelastic nematics of Kelvin-Voight type, considered by Warner, Terentjev and co-workers, when in (1) and ($2_3$), $\underline{\underline{e}}_p = 0$ and $\underline{\underline{\omega}}_p = 0$.

We analyze below only the nematic Maxwell case (i).

### *2.3. Viscoelastic nematics of Maxwell type*

In this case, $\underline{\underline{\sigma}}_p^s = \underline{\underline{\sigma}}_p^a = \underline{\underline{0}}$, so the relation ($7_3$) for entropy production takes the form:

$$TP_s = -\underline{q} \cdot \underline{\nabla} T + \underline{\underline{\sigma}}^s : \underline{\underline{e}}_p + \underline{\underline{\sigma}}^a : \underline{\underline{\omega}}_p . \qquad (7_4)$$

The symmetric $\underline{\underline{\sigma}}^s$ and asymmetric $\underline{\underline{\sigma}}_h^a$ parts of extra stress traceless tensor (the latter in the presence of magnetic field) are defined as:

$$\begin{aligned}\underline{\underline{\sigma}}^s = \partial f / \partial \underline{\underline{\varepsilon}} = G_0 \underline{\underline{\varepsilon}} + G_1[\underline{nn} \cdot \underline{\underline{\varepsilon}} + \underline{\underline{\varepsilon}} \cdot \underline{nn} - 2nn(\underline{\underline{\varepsilon}} : \underline{nn})] + 2(G_1 + G_2)(\underline{nn} - \underline{\underline{\delta}}/3)(\underline{\underline{\varepsilon}} : \underline{nn}) \\ + G_3(\underline{nn} \cdot \underline{\underline{\Omega}}_e - \underline{\underline{\Omega}}_e \cdot \underline{nn})\end{aligned} \qquad (8_1)$$

$$\underline{\underline{\sigma}}_h^a = \underline{\underline{h}}_n + \underline{\underline{\sigma}}^a, \quad \underline{\underline{\sigma}}^a = \partial f / \partial \underline{\underline{\Omega}}_e = -G_4(\underline{nn} \cdot \underline{\underline{\varepsilon}} - \underline{\underline{\varepsilon}} \cdot \underline{nn}) + G_5(\underline{nn} \cdot \underline{\underline{\Omega}}_e + \underline{\underline{\Omega}}_e \cdot \underline{nn}). \qquad (8_2)$$

Here the Onsager relation $G_4 \equiv -G_3$ automatically follows from (5) and definitions of $\underline{\underline{\sigma}}^s$ and $\underline{\underline{\sigma}}^a$ parts of extra stress tensor in ($8_{1,2}$).

When the inertial effects are negligible, the equilibrium equation for internal torques, independent of constitutive properties of asymmetric stress, has the form:

$$\underline{\underline{\sigma}}_h^a \cdot \underline{n} = 1/2[\underline{h} - \underline{n}(\underline{h}\cdot\underline{n})] \equiv \underline{h}^{\perp}/2. \qquad (8_3)$$

Substituting in this equation $\underline{\underline{h}}_n$ from ($6_2$) yields:

$$\underline{\underline{\sigma}}^a \cdot \underline{n} \equiv \underline{\Sigma} = \chi_a(\underline{n}\cdot\underline{H})[\underline{H} - \underline{n}(\underline{n}\cdot\underline{H})] = \underline{h}^{\perp}. \qquad (8_4)$$

Here $\underline{\underline{\sigma}}^a$ defined in ($8_2$), is the part of asymmetric stress tensor due to the viscoelastic effects. Note that the common ½ multiplier in the right-hand side of ($8_4$) is automatically changed here for the unity. This reflects the direct effect of magnetic field on the stress.

Due to ($7_4$), the constitutive relations between the irreversible kinematical variables and stress are of the LEP type:

$$\begin{aligned}\underline{\underline{\sigma}}^s = {} & \eta_0 \underline{\underline{e}}_p + \eta_1[\underline{\underline{nn}}\cdot\underline{\underline{e}}_p + \underline{\underline{e}}_p\cdot\underline{\underline{nn}} - 2\underline{\underline{nn}}(\underline{\underline{nn}}:\underline{\underline{e}}_p)] + 2(\eta_1+\eta_2)(\underline{\underline{nn}} - \underline{\underline{\delta}}/3)(\underline{\underline{nn}}:\underline{\underline{e}}_p) \\ & + \eta_3(\underline{\underline{nn}}\cdot\underline{\underline{\omega}}_p - \underline{\underline{\omega}}_p\cdot\underline{\underline{nn}})\end{aligned}, \qquad (9_1)$$

$$\underline{\underline{\sigma}}^a = -\eta_4(\underline{\underline{nn}}\cdot\underline{\underline{e}}_p - \underline{\underline{e}}_p\cdot\underline{\underline{nn}}) + \eta_5(\underline{\underline{nn}}\cdot\underline{\underline{\omega}}_p + \underline{\underline{\omega}}_p\cdot\underline{\underline{nn}}) \quad (\eta_4 = -\eta_3). \qquad (9_2)$$

Here we used the Onsager relation: $\eta_4 = -\eta_3$. It should be mentioned that the Onsager relations in the presence of magnetic field are: $\eta_4(\underline{H}) = -\eta_3(-\underline{H})$ [e.g. see de Groot and Mazur (1962), Landau and Lifshitz (1964)] Unlike the nematic ferrofluids, in case of LC's and LCP's, the dependence of kinetic coefficients on magnetic field is commonly ignored [see Lubensky (1973), Jarkova *et al* (2001)]. Therefore we consider the kinetic coefficients $\eta_k$, as well as the equilibrium parameters $G_k$, being independent of $\underline{H}$.

Finally, the anisotropic Fourier law that results from ($7_4$) has the form:

$$\underline{q} = -\underline{\underline{\kappa}}(\underline{n},T)\cdot\underline{\nabla}T; \quad \underline{\underline{\kappa}}(\underline{n},T) = \kappa_{\perp}(T)\underline{\underline{\delta}} + \kappa_a(T)\underline{\underline{nn}} \quad (\kappa_a = \kappa_{\parallel} - \kappa_{\perp}). \qquad (10)$$

As shown in Appendix, the $G_k$ and $\eta_k$ represent in this theory the scaling factors for independent nematic basis tensors. Therefore to avoid the degeneration of CE's we further assume that $G_k \neq 0$ and $\eta_k \neq 0$.

Demanding the thermodynamic stability for the elastic potential (5) results in the necessary and sufficient stability conditions, imposed on the parameters $G_k$, established by Leonov and Volkov (2004a),

$$G_0 > 0;\quad G_5 > 0;\quad G_0 + G_1 > 0;\quad 3/4G_0 + G_1 + G_2 > 0;\quad (G_0+G_1)G_5 > G_3^2. \qquad (11_1)$$

The thermodynamic stability conditions for dissipation in (8) in incompressible case, established by Leonov and Volkov (2004b), are the same as in ($11_1$) with substitution $G_k \to \eta_k$, i.e.

$$\eta_0 > 0;\ \eta_5 > 0;\ \eta_0 + \eta_1 > 0;\ 3/4\eta_0 + \eta_1 + \eta_2 > 0;\ (\eta_0 + \eta_1)\eta_5 > \eta_3^2 \qquad (11_2)$$

The thermodynamic stability conditions for the thermal processes are:

$$\kappa_\perp > 0, \quad \kappa_{\parallel} > 0\ (\kappa_\perp + \kappa_a > 0). \qquad (11_3)$$

Substituting now CE's ($9_{1,2}$) and (10) into the expression for entropy production ($7_4$) reduces the latter to the quadratic form:

$$TP_s = \underline{\underline{\kappa}}(\underline{n},T)\cdot\underline{\nabla T}\underline{\nabla T} + \eta_0\left|\underline{\underline{e}}_p\right|^2 + 2\eta_1\underline{nn}:\underline{\underline{e}}_p^2 + 2\eta_2(\underline{nn}:\underline{\underline{e}}_p)^2 - 4\eta_{3*}\underline{nn}:(\underline{\underline{e}}_p\cdot\underline{\underline{\omega}}_p) - 2\eta_5\underline{nn}:\underline{\underline{\omega}}_p^2, \quad (12)$$

which due to the stability constraints ($11_{1\text{-}3}$) is positively definite.

From thermodynamic viewpoint, the quantities $\underline{\underline{\omega}}_p$ and $\underline{\underline{e}}_p$ in (2)-(4) are not independent state variables but the measures of deviation of rates of "quasi-equilibrium" stated variables from their full kinematical parts. Therefore the variables $\underline{\underline{\omega}}_p$ and $\underline{\underline{e}}_p$ will be excluded from the final formulation of the theory, giving place to coupled equations for evolution of the "hidden" thermodynamic parameters $\underline{\underline{\Omega}}_e$ and $\underline{\underline{\varepsilon}}$.

## 3. Non-symmetric formulation of CE's using linear nematic operators

In order to derive the evolution equations for the hidden thermodynamic parameters $\underline{\underline{\varepsilon}}$ and $\underline{\underline{\Omega}}_e$, and an evolution equation for director $\underline{n}$, we use the theory of nematic operators (or N-operators) developed in Appendix. We distinguish below the second rank tensors characterizing the kinematical and dynamic properties of viscoelastic nematic liquid, from the third and fourth rank tensors (or *3-tensors* and *4-tensors*) that characterize the N-operators introduced in Appendix.

### *3.1. N-operator formulation*

We now represent the "elastic" and "viscous" CE's ($8_{1,2}$) and ($9_{1,2}$) in the compact operator form:

$$\underline{\underline{\hat{\sigma}}} \equiv \underline{\underline{\sigma}} - \underline{\underline{h}}_n = \mathbf{G}(\underline{n}) \bullet \underline{\underline{\Gamma}}_e = \sum_{k=0}^{5} \hat{G}_k \mathbf{a}_k(\underline{n}) \bullet \underline{\underline{\Gamma}}_e, \quad \underline{\underline{\hat{\sigma}}} = \boldsymbol{\eta}(\underline{n}) \bullet \underline{\underline{\gamma}}_p = \sum_{k=0}^{5} \hat{\eta}_k \mathbf{a}_k(\underline{n}) \bullet \underline{\underline{\gamma}}_p \qquad (13_{1,2})$$

Here $\mathbf{G}(\underline{n}) \equiv \mathbf{R}_G^o(\underline{n})$ and $\boldsymbol{\eta}(\underline{n}) \equiv \mathbf{R}_\eta^o(\underline{n})$ are the Onsager N-operators (or ON-operators) of moduli and viscosity represented by their 4-tensors, with the basis 4-tensors $\mathbf{a}_k(\underline{n})$ defined in (A1) in Appendix. The Onsager properties of CE's ($13_{1,2}$) are shown below in relations ($15_{1,2}$). Hereafter we use the convenient notation: $\underline{\underline{\hat{\sigma}}} \equiv \underline{\underline{\sigma}} - \underline{\underline{h}}_n$. Splitting equations ($13_{1,2}$) into symmetric and asymmetric parts yields:

$$\underline{\underline{\sigma}}^s = \sum_{k=0}^{2} \hat{G}_k \mathbf{a}_k(\underline{n}) \bullet \underline{\underline{\varepsilon}} + G_3 \mathbf{a}_3(\underline{n}) \bullet \underline{\underline{\Omega}}_e, \quad \underline{\underline{\sigma}}^a = G_4 \mathbf{a}_4(\underline{n}) \bullet \underline{\underline{\varepsilon}}_e + G_5 \mathbf{a}_5(\underline{n}) \bullet \underline{\underline{\Omega}}_e \qquad (14_1)$$

$$\underline{\underline{\sigma}}^s = \sum_{k=0}^{2} \hat{\eta}_k \mathbf{a}_k(\underline{n}) \bullet \underline{\underline{e}}_p + \eta_3 \mathbf{a}_3(\underline{n}) \bullet \underline{\underline{\omega}}_p, \quad \underline{\underline{\sigma}}^a = \eta_4 \mathbf{a}_4(\underline{n}) \bullet \underline{\underline{e}}_p + \eta_5 \mathbf{a}_5(\underline{n}) \bullet \underline{\underline{\omega}}_p \qquad (14_2)$$

The following notations have been introduced in ($13_{1,2}$) and ($14_{1,2}$):

$$\hat{G}_k = G_k \ (k = 0,1,3,4,5), \ \ \hat{G}_2 = 2G_1 + 2G_2, \ G_4 = -G_3 \qquad (15_1)$$

$$\hat{\eta}_k = \eta_k \ (k = 0,1,3,4,5), \ \ \hat{\eta}_2 = 2\eta_1 + 2\eta_2, \ \eta_4 = -\eta_3 \qquad (15_2)$$

The last equalities in ($15_1$) and ($15_2$) clearly demonstrate the Onsager relations between the tensor and pseudo-tensor variables in the CE's (13-$14_{1,2}$).

The N-operator presentation for free energy has the form:

$$f - f_m = 1/2 \sum_{k=0}^{5} \hat{G}_k \underline{\underline{\Gamma}}_e \bullet \mathbf{a}_k(\underline{n}) \bullet \underline{\underline{\Gamma}}_e = 1/2 \sum_{k=0}^{2} G_k \underline{\underline{\varepsilon}} \bullet \mathbf{a}_k(\underline{n}) \bullet \underline{\underline{\varepsilon}} + G_3 \underline{\underline{\varepsilon}} \bullet \mathbf{a}_3(\underline{n}) \bullet \underline{\underline{\Omega}}_e + 1/2 G_5 \underline{\underline{\Omega}}_e \bullet \mathbf{a}_5(\underline{n}) \bullet \underline{\underline{\Omega}}_e \quad (16_1)$$

The expression for dissipation can also be presented in the N-operator form:

$$D \equiv TP_s|_T = \sum_{k=0}^{5} \hat{\eta}_k \underline{\underline{\gamma}}_p \bullet \mathbf{a}_k(\underline{n}) \bullet \underline{\underline{\gamma}}_p = \sum_{k=0}^{2} \hat{\eta}_k \underline{\underline{e}}_p \bullet \mathbf{a}_k(\underline{n}) \bullet \underline{\underline{e}}_p + 2\eta_3 \underline{\underline{e}}_p \bullet \mathbf{a}_3(\underline{n}) \bullet \underline{\underline{\omega}}_p + \eta_5 \underline{\underline{\omega}}_p \bullet \mathbf{a}_5(\underline{n}) \bullet \underline{\underline{\omega}}_p \quad (16_2)$$

The thermodynamic stability requirements guaranty the existence of inverse operations for ($13_{1,2}$), ($14_{1,2}$). Because the inverse ON-operators belong to the ON class, we present the inverse relations for equations ($13_1$) and ($14_1$) as:

$$\begin{gathered} \underline{\underline{\Gamma}}_e = \mathbf{G}^{-1}(\underline{n}) \bullet \underline{\underline{\hat{\sigma}}} \equiv \mathbf{J}(\underline{n}) \bullet \underline{\underline{\hat{\sigma}}}, \quad \mathbf{J}(\underline{n}) = \sum_{k=0}^{5} J_k \mathbf{a}_k(\underline{n}) \\ \underline{\underline{\varepsilon}} = \sum_{k=0}^{2} J_k \mathbf{a}_k(\underline{n}) \bullet \underline{\underline{\sigma}}^s + J_3 \mathbf{a}_3(\underline{n}) \bullet \underline{\underline{\sigma}}^a, \quad \underline{\underline{\Omega}}_e = J_4 \mathbf{a}_4(\underline{n}) \bullet \underline{\underline{\sigma}}^s + J_5 \mathbf{a}_5(\underline{n}) \bullet \underline{\underline{\sigma}}^a \end{gathered} . \qquad (17_1)$$

Similarly, the inverse relations for equations ($13_2$) and ($14_2$) are presented as:

$$\underset{=p}{\gamma} = \boldsymbol{\eta}^{-1}(\underline{n})\bullet\underline{\underline{\hat{\sigma}}} \equiv \boldsymbol{\Phi}(\underline{n})\bullet\underline{\underline{\hat{\sigma}}}, \quad \boldsymbol{\Phi}(\underline{n}) = \sum_{k=0}^{5}\varphi_k\mathbf{a}_k(\underline{n})$$

$$\underset{=p}{e} = \sum_{k=0}^{2}\varphi_k\mathbf{a}_k(\underline{n})\bullet\underline{\underline{\sigma}}^s + \varphi_3\mathbf{a}_3(\underline{n})\bullet\underline{\underline{\sigma}}^a, \quad \underset{=p}{\omega} = \varphi_4\mathbf{a}_4(\underline{n})\bullet\underline{\underline{\sigma}}^s + \varphi_5\mathbf{a}_5(\underline{n})\bullet\underline{\underline{\sigma}}^a \qquad (17_2)$$

Here $\mathbf{J}(\underline{n}) \equiv \mathbf{R}^o_J(\underline{n})$ and $\boldsymbol{\Phi}(\underline{n}) \equiv \mathbf{R}^o_\varphi(\underline{n})$ are the compliance and fluidity ON-operators, respectively. Their basis scalars, compliance $J_k$ (dimensionality of inverse modulus) in ($17_1$), and "fluidity" $\varphi_k$ (dimensionality of inverse viscosity) in ($17_2$), are presented with the use of formulae(A.$12_2$): for parameters of inverse ON-operations, as:

$$J_0 = \frac{1}{G_0}, \; J_1 = \frac{(G_3^2 - G_1G_5)/G_0}{G_5(G_0+G_1)-G_3^2}, \; J_2 = -\frac{3/2(G_1+G_2)/G_0}{3/4G_0+G_1+G_2}$$
$$J_3 = -J_4 = \frac{-G_3}{G_5(G_0+G_1)-G_3^2}, \; J_5 = \frac{G_0+G_1}{G_5(G_0+G_1)-G_3^2}, \qquad (18_1)$$

$$\varphi_0 = \frac{1}{\eta_0}, \; \varphi_1 = \frac{(\eta_3^2 - \eta_1\eta_5)/\eta_0}{\eta_5(\eta_0+\eta_1)-\eta_3^2}, \; \varphi_2 = -\frac{3/2(\eta_1+\eta_2)/\eta_0}{3/4\eta_0+\eta_1+\eta_2} \qquad (18_2)$$
$$\varphi_3 = \frac{-\eta_3}{\eta_5(\eta_0+\eta_1)-\eta_3^2}, \; \varphi_4 = -\varphi_3, \; \varphi_5 = \frac{\eta_0+\eta_1}{\eta_5(\eta_0+\eta_1)-\eta_3^2}$$

One can see a similarity between ($18_1$) and ($18_2$). It is also seen that the thermodynamic stability conditions ($11_{1,2}$) guaranty the existence of inverse N-operations ($17_{1,2}$).

Using CE's ($13_{1,2}$) expresses $\underset{=p}{\gamma} = \underset{=p}{e} + \underset{=p}{\omega}$ via $\underset{=e}{\Gamma} = \underset{=e}{\varepsilon} + \underset{=e}{\Omega}$ and vice versa, as:

$$\underset{=p}{\gamma} = \mathbf{s}(\underline{n})\bullet\underset{=e}{\Gamma} \quad \left(\mathbf{s}(\underline{n}) = \sum_{k=0}^{5}s_k\mathbf{a}_k(\underline{n})\right) \; ; \quad \underset{=e}{\Gamma} = \boldsymbol{\theta}(\underline{n})\bullet\underset{=p}{\gamma} \quad \left(\boldsymbol{\theta}(\underline{n}) = \sum_{k=0}^{5}\theta_k\mathbf{a}_k(\underline{n})\right) \qquad (19_{1,2})$$

Here $\mathbf{s}(\underline{n}) = \boldsymbol{\eta}^{-1}(\underline{n})\bullet\mathbf{G}(\underline{n}) = \boldsymbol{\Phi}(\underline{n})\bullet\mathbf{G}(\underline{n})$ and $\boldsymbol{\theta}(\underline{n}) = \mathbf{G}^{-1}(\underline{n})\bullet\boldsymbol{\eta}(\underline{n}) = \mathbf{J}(\underline{n})\bullet\boldsymbol{\eta}(\underline{n})$ are the N-operators of relaxation frequencies and relaxation times, respectively. Using ($18_{1,2}$), their basis scalar parameters $s_k$ and $\theta_k$ are calculated as:

$$s_0 = \frac{G_0}{\eta_0}, \; s_1 = \frac{\eta_5(G_1\eta_0 - G_0\eta_1) + \eta_3(G_0\eta_3 - G_3\eta_0)}{\eta_0[\eta_5(\eta_0+\eta_1)-\eta_3^2]}, \; s_2 = \frac{3}{2}\cdot\frac{(G_1+G_2)\eta_0 - G_0(\eta_1+\eta_2)}{\eta_0(3/4\eta_0+\eta_1+\eta_2)} \qquad (20_1)$$
$$s_3 = \frac{G_3(\eta_0+\eta_1) - \eta_3(G_0+G_1)}{\eta_5(\eta_0+\eta_1)-\eta_3^2}, \; s_4 = \frac{G_5\eta_3 - G_3\eta_5}{\eta_5(\eta_0+\eta_1)-\eta_3^2}, \; s_5 = \frac{G_5(\eta_0+\eta_1) - G_3\eta_5}{\eta_5(\eta_0+\eta_1)-\eta_3^2}$$

$$\theta_0 = \frac{\eta_0}{G_0},\ \ \theta_1 = \frac{G_5(\eta_1 G_0 - \eta_0 G_1) + G_3(\eta_0 G_3 - \eta_3 G_0)}{G_0[G_5(G_0+G_1) - G_3^2]},\ \ \theta_2 = \frac{3}{2}\cdot\frac{(\eta_1+\eta_2)G_0 - \eta_0(G_1+G_2)}{G_0(3/4G_0 + G_1 + G_2)}$$

$$\theta_3 = \frac{\eta_3(G_0+G_1) - G_3(\eta_0+\eta_1)}{G_5(G_0+G_1) - G_3^2},\ \ \theta_4 = \frac{\eta_5 G_3 - \eta_3 G_5}{G_5(G_0+G_1) - G_3^2},\ \ \theta_5 = \frac{\eta_5(G_0+G_1) - \eta_3 G_5}{G_5(G_0+G_1) - G_3^2} \qquad (20_2)$$

Note that $\boldsymbol{\theta}(\underline{n}), \mathbf{s}(\underline{n}) \in \check{\mathrm{R}}_6$ are not Onsager N-operators, because their basis scalars have no Onsager symmetry. It is also easy to derive the relations:

$$\boldsymbol{\theta}(\underline{n}) = \boldsymbol{\eta}(\underline{n})\bullet\mathbf{G}^{-1}(\underline{n}),\ \ \mathbf{s}(\underline{n}) = \mathbf{G}(\underline{n})\bullet\boldsymbol{\eta}^{-1}(\underline{n}) = \boldsymbol{\theta}^{-1}(\underline{n}). \qquad (21)$$

Formulae (21) are anisotropic, non-symmetric N-operator analogs of the common scalar relations in isotropic case.

### *3.2. Evolution equations in non-symmetric case*

(i) *Evolution equations for elastic (transient) strain* $\underline{\underline{\varepsilon}}$ *and elastic rotation* $\underline{\underline{\Omega}}_e$

Substituting ($19_1$) in (4) yields the compact operator form of evolution equation for $\underline{\underline{\Gamma}}_e$:

$$\overset{0}{\underline{\underline{\Gamma}}}_e + \mathbf{s}(\underline{n})\bullet\underline{\underline{\Gamma}}_e = \underline{\underline{\gamma}}\ ,\quad \mathbf{s}(\underline{n}) \in \check{\mathrm{R}}_6 \qquad (22_1)$$

Here the basis scalars of relaxation frequency N-operator $\mathbf{s}(\underline{n})$ are presented in ($20_1$). Equation ($22_1$) written for symmetric and anti-symmetric terms is:

$$\overset{0}{\underline{\underline{\varepsilon}}} + \sum_{k=0}^{2} s_k \mathbf{a}_k(\underline{n})\bullet\underline{\underline{\varepsilon}} + s_3\mathbf{a}_3(\underline{n})\bullet\underline{\underline{\Omega}}_e = \underline{\underline{e}},\ \ \overset{0}{\underline{\underline{\Omega}}}_e + s_4\mathbf{a}_4(\underline{n})\bullet\underline{\underline{\varepsilon}} + s_5\mathbf{a}_5(\underline{n})\bullet\underline{\underline{\Omega}}_e = \underline{\underline{\omega}}. \qquad (22_2)$$

The evolution equation ($22_{1,2}$) written in the common tensor form is:

$$\overset{0}{\underline{\underline{\varepsilon}}} + s_0\underline{\underline{\varepsilon}} + s_1[\underline{nn}\cdot\underline{\underline{\varepsilon}} + \underline{\underline{\varepsilon}}\cdot\underline{nn} - 2nn(\underline{\underline{\varepsilon}}:\underline{nn})] + s_2(\underline{nn} - \underline{\underline{\delta}}/3)(\underline{\underline{\varepsilon}}:\underline{nn})$$
$$+ s_3(\underline{nn}\cdot\underline{\underline{\Omega}}_e - \underline{\underline{\Omega}}_e\cdot\underline{nn}) = \underline{\underline{e}} \qquad (22_3)$$
$$\overset{0}{\underline{\underline{\Omega}}}_e + s_4(\underline{nn}\cdot\underline{\underline{\varepsilon}} - \underline{\underline{\varepsilon}}\cdot\underline{nn}) + s_5(\underline{nn}\cdot\underline{\underline{\Omega}}_e + \underline{\underline{\Omega}}_e\cdot\underline{nn}) = \underline{\underline{\omega}}$$

Equations ($6_{1,2}$) and ($22_3$) constitute the closed set of CE's if parameter $\underline{n}$ is known. Note that the evolution equations (22) hold for non-isothermal situation.

(ii) *Evolution equations for stress*

We now consider isothermal situations and using the Jaumann differentiation of terms in CE's will treat the director $\underline{n}$ as a "frozen" variable. This is because in the weak nematic viscoelasticity, expression of $\overset{0}{\underline{n}}$ via $\underline{\underline{\varepsilon}}$ involves the negligible terms $O(\underline{\underline{\varepsilon}}^2)$. Substituting then the expressions $\underline{\underline{\Gamma}}_e = \mathbf{J}(\underline{n})\bullet\underline{\underline{\hat{\sigma}}}$, $\underline{\underline{\gamma}}_p = \mathbf{\Phi}(\underline{n})\bullet\underline{\underline{\hat{\sigma}}}$ from ($17_{1,2}$) into kinematical relation (4) yields the following equivalent N-operator forms of the nematic Maxwell equation:

$$\mathbf{J}(\underline{n})\bullet\overset{0}{\underline{\underline{\hat{\sigma}}}} + \mathbf{\Phi}(\underline{n})\bullet\underline{\underline{\hat{\sigma}}} = \overset{0}{\underline{\underline{\gamma}}},\quad \overset{0}{\underline{\underline{\hat{\sigma}}}} + \mathbf{s}(\underline{n})\bullet\underline{\underline{\hat{\sigma}}} = \mathbf{G}(\underline{n})\bullet\overset{0}{\underline{\underline{\gamma}}},\quad \boldsymbol{\theta}(\underline{n})\bullet\overset{0}{\underline{\underline{\hat{\sigma}}}} + \underline{\underline{\hat{\sigma}}} = \boldsymbol{\eta}(\underline{n})\bullet\overset{0}{\underline{\underline{\gamma}}} \qquad (23_{1,2,3})$$

The basis scalar parameters $J_k, \varphi_k$ and $s_k, \theta_k$ were expressed via the given model parameters $G_k$ and $\eta_k$ in ($18_{1,2}$) and ($20_{,2}$), respectively. The CE's ($23_{1,2,3}$) can also be presented in the form with splitting symmetric and anti-symmetric parts. For example such an expression of ($23_2$) is:

$$\begin{aligned} &\overset{0}{\underline{\underline{\sigma}}}{}^s + \sum_{k=0}^{2} \mathbf{a}_k(\underline{n})\bullet(s_k\underline{\underline{\sigma}}^s - G_k\underline{\underline{e}}) + \mathbf{a}_3(\underline{n})\bullet(s_3\underline{\underline{\sigma}}^a - G_3\underline{\underline{\omega}}) = \underline{\underline{0}} \\ &\overset{0}{\underline{\underline{\sigma}}}{}^a + \mathbf{a}_4(\underline{n})\bullet(s_4\underline{\underline{\sigma}}^s - G_3\underline{\underline{e}}) + \mathbf{a}_5(\underline{n})\bullet(s_5\underline{\underline{\sigma}}^a - G_5\underline{\underline{\omega}}) = \underline{\underline{0}} \end{aligned} \qquad (23_2^*)$$

We avoid here writing the evident but awkward tensor presentation for equation ($23_2^*$).

(iii) *Evolution equation for director*

Left multiplying by $\mathbf{b}(\underline{n})$ two matrix equations ($22_2$) and the last equation ($14_2$), with the use of formulae (A4) and (A5), and treating $\underline{n}$ as a frozen variable, results in:

$$\begin{gathered} \overset{0}{\underline{P}} + (s_0 + s_1)\underline{P} - s_3\underline{Q} = \mathbf{b}(\underline{n})\bullet\underline{\underline{e}},\quad \overset{0}{\underline{Q}} + s_4\underline{Q} - s_3\underline{P} = \overset{0}{\underline{n}},\quad -G_3\underline{P} + G_4\underline{Q} = \underline{\Sigma} \\ (\underline{P} \equiv \mathbf{b}(\underline{n})\bullet\underline{\underline{\varepsilon}},\quad \underline{Q} \equiv \underline{\underline{\Omega}}_e\cdot\underline{n},\quad \underline{\Sigma} = \underline{\underline{\sigma}}^a\cdot\underline{n}) \end{gathered} \qquad (24)$$

Excluding $\underline{P}, \underline{Q}$ from the first two equations in (24) with the aid of the third one yields:

$$G_4\left(\overset{00}{\underline{n}} + \underline{n}\left|\overset{0}{\underline{n}}\right|^2 + \alpha\overset{0}{\underline{n}}\right) - G_3\mathbf{b}(\underline{n})\bullet(\overset{0}{\underline{\underline{e}}} + \beta\underline{\underline{e}}) = \overset{00}{\underline{\Sigma}} + \underline{n}\cdot\left|\overset{0}{\underline{\Sigma}}\right|^2 + \gamma\overset{0}{\underline{\Sigma}} + \delta\underline{\Sigma}. \qquad (25)$$

As seen, the scalar multiplication of (25) by $\underline{n}$ nullifies both sides of this equation. The parameters in (25) are expressed via parameters in (24) as:

$$\alpha = s_0 + s_1 - s_3G_3/G_5,\quad \beta = s_5 + s_4G_5/G_3,\quad \gamma = \alpha + s_4 - s_3G_3/G_5,\quad \delta = \alpha s_5 - \beta s_3G_3/G_5.$$

Equation (25) is not closed unless the force $\underline{\Sigma} = \underline{\underline{\sigma}}^a\cdot\underline{n}$ in (24) is specified. In case of LCP, several physical sources cause possible occurrence of this force: the presence of magnetic (or electric) external fields, the inertial effects of internal rotations, the Cosserat effects of internal couples, and the Frank elasticity. Except for the first one, all other effects seem to be negligible for LCP's. Considering only the effect of magnetic field, results in equation ($8_4$): $\underline{\Sigma} = \underline{h}^{\perp}$, where $\underline{h}^{\perp}$ is the transverse component of the molecular field. Then, substituting ($8_4$) into (25) closes the evolution equation for director.

## 4. Symmetric theory

In the absence of magnetic field ($\underline{H} = 0$), $\underline{\Sigma} = \underline{\underline{\sigma}}^a \cdot \underline{n} = 0$, so stress tensor is symmetric.

### *4.1. Tensor formulation of symmetric theory*

In symmetric case when $\underline{\underline{\sigma}}^a = 0$, equations ($8_2$), ($9_2$) yield the kinematical relations:

$$\underline{\underline{\Omega}}_e = \lambda_1(\underline{\underline{\varepsilon}} \cdot \underline{nn} - \underline{nn} \cdot \underline{\underline{\varepsilon}}), \quad \underline{\underline{\omega}}_p = \lambda_2(\underline{\underline{e}}_p \cdot \underline{nn} - \underline{nn} \cdot \underline{\underline{e}}_p) \qquad (\lambda_1 = G_3/G_5,\ \lambda_2 = \eta_3/\eta_5) \tag{$26_{1,2}$}$$

where $\lambda_1$ and $\lambda_2$ are sign indefinite parameters. Substituting ($26_{1,2}$) back into ($8_2$), ($9_2$), and in (5) yields the reduced (or "renormalized") formulation of these equations with symmetric stress as:

$$f = 1/2G_0\left|\underline{\underline{\varepsilon}}\right|^2 + G_1^r \underline{nn} : \underline{\underline{\varepsilon}}^2 + G_2^r(\underline{nn} : \underline{\underline{\varepsilon}})^2 \quad (G_1^r = G_1 - G_3^2/G_4,\ G_2^r = G_2 + G_3^2/G_4) \tag{27}$$

$$\underline{\underline{\sigma}} = \partial f/\partial \underline{\underline{\varepsilon}} = G_0 \underline{\underline{\varepsilon}} + G_1^r[\underline{nn} \cdot \underline{\underline{\varepsilon}} + \underline{\underline{\varepsilon}} \cdot \underline{nn} - 2\underline{nn}(\underline{\underline{\varepsilon}} : \underline{nn})] + 2(G_1^r + G_2^r)(\underline{\underline{\varepsilon}} : \underline{nn})(\underline{nn} - 1/3\underline{\underline{\delta}}) \tag{$28_1$}$$

$$\underline{\underline{\sigma}} = \frac{1}{2}\frac{\partial D}{\partial \underline{\underline{e}}_p} = \eta_0 \underline{\underline{e}}_p + \eta_1^r[\underline{nn} \cdot \underline{\underline{e}}_p + \underline{\underline{e}}_p \cdot \underline{nn} - 2\underline{nn}(\underline{\underline{e}}_p : \underline{nn})] + 2(\eta_1^r + \eta_2^r)(\underline{\underline{e}}_p : \underline{nn})(\underline{nn} - \underline{\underline{\delta}}/3) \tag{$28_2$}$$

Additionally, the dissipation $D = TP_s\big|_T = \underline{\underline{\sigma}} : \underline{\underline{e}}_p$ is presented in the form:

$$D/2 = \eta_0\left|\underline{\underline{e}}_p\right|^2 + \eta_1^r \underline{nn} : \underline{\underline{e}}_p^2 + \eta_2^r(\underline{nn} : \underline{\underline{e}}_p)^2 \quad (\eta_1^r = \eta_1 - \eta_3^2/\eta_4,\ \eta_2^r = \eta_2 + \eta_3^2/\eta_4) \tag{29}$$

It is seen that D/2 plays the role of Raleigh (potential) dissipative function. For simplicity we change hereafter the notations as: $G_k^r \leftrightarrow G_k$, $\eta_k^r \leftrightarrow \eta_k$ $(k = 1,2)$.

The (necessary and sufficient) conditions of *thermodynamic stability* ($11_{1,2}$) are:

$$G_0 > 0,\ G_0 + G_1 > 0,\ 3/4G_0 + G_1 + G_2 > 0 \tag{$30_1$}$$

$$\eta_0 > 0,\ \eta_0 + \eta_1 > 0,\ 3/4\eta_0 + \eta_1 + \eta_2 > 0. \tag{$30_2$}$$

One can see the complete similarity in inequalities ($30_{1,2}$).

### *4.2. N-operator formulation of symmetric theory*

Using Appendix, kinematical relations ($26_{1,2}$) are represented in the operator form:

$$\overset{0}{\underline{\underline{\Omega}}}_e \cdot \underline{n} = \lambda_1 \mathbf{b}(\underline{n}) \bullet \underline{\underline{\varepsilon}}, \quad \overset{0}{\underline{\underline{\omega}}}_p \cdot \underline{n} = \lambda_2 \mathbf{b}(\underline{n}) \bullet \underline{\underline{e}}_p . \qquad (31_{1,2})$$

Here the Jaumann differentiation has been performed with "frozen" director.

N-operator presentation of CE's ($30_{1,2}$) is:

$$\underline{\underline{\sigma}} = \mathbf{G}(\underline{n}) \bullet \underline{\underline{\varepsilon}} = \boldsymbol{\eta}(\underline{n}) \bullet \underline{\underline{e}}_p ; \quad \mathbf{G}(\underline{n}) = \sum_{k=0}^{2} \hat{G}_k \mathbf{a}_k(\underline{n}), \quad \boldsymbol{\eta}(\underline{n}) = \sum_{k=0}^{2} \hat{\eta}_k \mathbf{a}_k(\underline{n}) \qquad (32_{1,2})$$

Here the basis numerical 4-tensors $\mathbf{a}_s(\underline{n})$ ($s = 0,1,2$) and 3-tensor for the operator $\mathbf{b}(\underline{n})$ are shown in (A1) of Appendix. Parameters $\hat{G}_k$ and $\hat{\eta}_k$ are represented via $G_k$ and $\eta_k$ as:

$$\hat{G}_0 = G_0, \; \hat{G}_1 = G_1, \; \hat{G}_2 = 2G_1 + 2G_2 ; \quad \hat{\eta}_0 = \eta_0, \; \hat{\eta}_1 = \eta_1, \; \hat{\eta}_2 = 2\eta_1 + 2\eta_2 . \qquad (33_{1,2})$$

The free energy (27) and dissipation (29) are expressed in the N-operator form as:

$$f = 1/2 \underline{\underline{\varepsilon}} \bullet \mathbf{G}(\underline{n}) \bullet \underline{\underline{\varepsilon}} , \quad D = \underline{\underline{e}}_p \bullet \boldsymbol{\eta}(\underline{n}) \bullet \underline{\underline{e}}_p . \qquad (34_{1,2})$$

Due to the stability constraints ($30_{1,2}$) the symmetric N-operators $\mathbf{G}(\underline{n})$ and $\boldsymbol{\eta}(\underline{n})$ are positively definite, so the reciprocal $\mathbf{G}^{-1}(\underline{n})$ and $\boldsymbol{\eta}^{-1}(\underline{n})$ exist and are positively definite too. As shown in Appendix, it means that $\mathbf{G}(\underline{n}), \boldsymbol{\eta}(\underline{n}) \in \check{S}_3$, where $\check{S}_3$ is the three parametric, transversally isotropic (TI), commutative subgroup of N-operators. Using properties of this subgroup, we can present the inverse TI-operators as:

$$\mathbf{G}^{-1}(\underline{n}) \equiv \mathbf{J}(\underline{n}) = \sum_{i=0}^{3} J_k \mathbf{a}_k(\underline{n}) , \qquad \boldsymbol{\eta}^{-1}(\underline{n}) \equiv \boldsymbol{\Phi}(\underline{n}) = \sum_{i=0}^{3} \varphi_k \mathbf{a}_k(\underline{n}) . \qquad (35_{1,2})$$

Here $\mathbf{J}(\underline{n})$ and $\boldsymbol{\Phi}(\underline{n})$ are the compliance and fluidity TI-operators, respectively. Due to (A18) the expressions for their respective basis scalars are:

$$J_0 = 1/G_0, \quad J_1 = -\frac{G_1/G_0}{G_0 + G_1}, \quad J_2 = -\frac{3/2(G_1 + G_2)/G_0}{3/4G_0 + G_1 + G_2} \qquad (36_1)$$

$$\varphi_0 = 1/\eta_0, \quad \varphi_1 = -\frac{\eta_1/\eta_0}{\eta_0 + \eta_1}, \quad \varphi_2 = -\frac{3/2(\eta_1 + \eta_2)/\eta_0}{3/4\eta_0 + \eta_1 + \eta_2} . \qquad (36_2)$$

One can see that the TI-operators $\boldsymbol{\eta}^{-1}(\underline{n})$ and $\mathbf{G}^{-1}(\underline{n})$ exist if the thermodynamic stability conditions ($30_{1,2}$) are satisfied.

Using the first ("dual") equation in ($32_1$), we can express $\underline{\underline{e}}_p$ via $\underline{\underline{\varepsilon}}_e$ or *vice versa:*

$$\underline{\underline{e}}_p = \boldsymbol{\theta}(\underline{n})\bullet\underline{\underline{\varepsilon}}, \qquad \underline{\underline{\varepsilon}} = \mathbf{s}(\underline{n})\bullet\underline{\underline{e}}_p \ . \tag{$37_{1.2}$}$$

Here the TI-operators of relaxation times $\boldsymbol{\theta}(\underline{n})$ and relaxation frequency $\mathbf{s}(\underline{n}) \equiv \boldsymbol{\theta}^{-1}(\underline{n})$ are:

$$\boldsymbol{\theta}(\underline{n}) = \boldsymbol{\eta}(\underline{n})\bullet\mathbf{G}^{-1}(\underline{n}) = \sum_{i=0}^{2}\theta_i\mathbf{a}_i(\underline{n}), \quad \mathbf{s}(\underline{n}) = \mathbf{G}(\underline{n})\bullet\boldsymbol{\eta}^{-1}(\underline{n}) = \sum_{i=0}^{2}s_i\mathbf{a}_i(\underline{n}). \tag{$38_{1.2}$}$$

Using ($36_{1,2}$) the basis scalar parameters $\theta_i$ and $s_i$ are calculated as:

$$\theta_0 = \frac{\eta_0}{G_0}, \quad \theta_1 = \frac{\eta_1 G_0 - G_1\eta_0}{G_0(G_0+G_1)}, \quad \theta_2 = \frac{3}{2}\cdot\frac{G_0(\eta_1+\eta_2)-\eta_0(G_1+G_2)}{G_0(3/4G_0+G_1+G_2)} \tag{$39_1$}$$

$$s_0 = \frac{G_0}{\eta_0}, \quad s_1 = \frac{G_1\eta_0 - \eta_1 G_0}{\eta_0(\eta_0+\eta_1)}, \quad s_2 = \frac{3}{2}\cdot\frac{\eta_0(G_1+G_2)-G_0(\eta_1+\eta_2)}{\eta_0(3/4\eta_0+\eta_1+\eta_2)}. \tag{$39_2$}$$

It is seen that formulae in ($39_1$) change for those in ($39_2$) and *vice versa* when the substitutions $G_k \to \eta_k \to G_k$ hold simultaneously. Note that in spite of positive definiteness of all the 4-tensors, characterizing elastic, viscous and relaxation properties, not all their basic scalars are necessarily positive.

### *4.3. Evolution equations in symmetric case*

(i) *Evolution equations for elastic transient strain tensor* $\underline{\underline{\varepsilon}}$

Substituting ($37_1$) into kinematical equation (1) yields the TI-operator form of evolution equation for transient elastic strain $\underline{\underline{\varepsilon}}$:

$$\overset{0}{\underline{\underline{\varepsilon}}} + \mathbf{s}(\underline{n})\bullet\underline{\underline{\varepsilon}} = \underline{\underline{e}}\ , \quad \mathbf{s}(\underline{n}) = \sum_{i=0}^{2}s_i\mathbf{a}_i(\underline{n}) \in \check{S}_3. \tag{$40_1$}$$

The evolution equation ($40_1$) written in the common tensor form is:

$$\overset{0}{\underline{\underline{\varepsilon}}} + s_0\underline{\underline{\varepsilon}} + s_1[\underline{\underline{nn}}\cdot\underline{\underline{\varepsilon}} + \underline{\underline{\varepsilon}}\cdot\underline{\underline{nn}} - 2nn(\underline{\underline{\varepsilon}}:\underline{\underline{nn}})] + s_2(\underline{\underline{nn}} - \underline{\underline{\delta}}/3)(\underline{\underline{\varepsilon}}:\underline{\underline{nn}}) = \underline{\underline{e}}\ . \tag{$40_2$}$$

Here parameters $s_k$ are expressed in ($39_2$) via the basic parameters $G_k$ and $\eta_k$ of nematic Maxwell model.

(ii) *Evolution equations for stress*

We remind once again that when using Jaumann differentiation of terms in CE's, the director $\underline{n}$ is treated as the "frozen" variable. Then substituting the TI-operator expressions $\varepsilon = \mathbf{J}(\underline{n}) \bullet \underline{\underline{\sigma}}$ and $\underline{\underline{e}}_p = \mathbf{\Phi}(\underline{n}) \bullet \underline{\underline{\sigma}}$ into kinematical relation (1) yields the following equivalent operator forms for nematic Maxwell equation:

$$\mathbf{J}(\underline{n}) \bullet \overset{0}{\underline{\underline{\sigma}}} + \mathbf{\Phi}(\underline{n}) \bullet \underline{\underline{\sigma}} = \underline{\underline{e}}, \quad \overset{0}{\underline{\underline{\sigma}}} + \mathbf{s}(\underline{n}) \bullet \underline{\underline{\sigma}} = \mathbf{G}(\underline{n}) \bullet \underline{\underline{e}}, \quad \mathbf{\theta}(\underline{n}) \bullet \overset{0}{\underline{\underline{\sigma}}} + \underline{\underline{\sigma}} = \mathbf{\eta}(\underline{n}) \bullet \underline{\underline{e}} . \qquad (41_{1,2,3})$$

Here $G_k$ and $\eta_k$ are given model parameters, the parameters $J_k, \varphi_k$ are presented in $(36_{1,2})$ and $s_k, \theta_k$ in $(39_{1,2})$, via $G_k$ and $\eta_k$. These CE's can be written in the more detailed matrix form. For example, the relation $(41_2)$ is presented as:

$$\overset{0}{\underline{\underline{\sigma}}} + \sum_{k=0}^{2} \mathbf{a}_k(\underline{n}) \bullet (s_k \underline{\underline{\sigma}} - G_k \underline{\underline{e}}) = \underline{\underline{0}}. \qquad (42_1)$$

Equation $(42_1)$ written in the common tensor form is:

$$\begin{aligned} &\overset{0}{\underline{\underline{\sigma}}} + s_0 \underline{\underline{\sigma}} + s_1[\underline{nn} \cdot \underline{\underline{\sigma}} + \underline{\underline{\sigma}} \cdot \underline{nn} - 2nn(\underline{\underline{\sigma}} : \underline{nn})] + s_2(\underline{nn} - \underline{\underline{\delta}}/3)(\underline{\underline{\sigma}} : \underline{nn}) \\ &= G_0 \underline{\underline{e}} + G_1[\underline{nn} \cdot \underline{\underline{e}} + \underline{\underline{e}} \cdot \underline{nn} - 2nn(\underline{\underline{e}} : \underline{nn})] + G_2(\underline{nn} - \underline{\underline{\delta}}/3)(\underline{\underline{e}} : \underline{nn}) \end{aligned} \qquad (42_2)$$

(iii) *Evolution equation for director*

Using the operations with $\mathbf{b}(\underline{n})$ introduced in Appendix, it is easy to obtain from the (1)-(3) the coupled relations:

$$\overset{0}{\underline{n}} = \mathbf{b}(\underline{n}) \bullet (\lambda_1 \overset{0}{\underline{\underline{\varepsilon}}} + \lambda_2 \underline{\underline{e}}_p), \quad \mathbf{b}(\underline{n}) \bullet \underline{\underline{e}} = \mathbf{b}(\underline{n}) \bullet (\overset{0}{\underline{\underline{\varepsilon}}} + \underline{\underline{e}}_p),$$

where the parameters $\lambda_1$ and $\lambda_2$ are defined in (26). Substituting here $(37_1)$ with the aid of (A4) and (A5) yields:

$$\theta^* \overset{0}{\underline{n}} = \mathbf{b}(\underline{n}) \bullet (\lambda_1 \theta^* \overset{0}{\underline{\underline{\varepsilon}}} + \lambda_2 \underline{\underline{\varepsilon}}), \quad \theta^* \mathbf{b}(\underline{n}) \bullet \underline{\underline{e}} = \mathbf{b}(\underline{n}) \bullet (\theta^* \overset{0}{\underline{\underline{\varepsilon}}} + \underline{\underline{\varepsilon}}).$$

Excluding $\mathbf{b}(\underline{n}) \bullet \underline{\underline{\varepsilon}}$ from these coupled equations yields the evolution equation for director:

$$\theta^* \left( \overset{00}{\underline{n}} + \underline{n} \cdot \left| \overset{0}{\underline{n}} \right|^2 \right) + \overset{0}{\underline{n}} = \mathbf{b}(\underline{n}) \bullet (\lambda_1 \theta^* \overset{0}{\underline{\underline{e}}} + \lambda_2 \underline{\underline{e}}). \qquad (43)$$

$$\theta^* = (s_1 + s_2)^{-1} = (\eta_0 + \eta_1)/(G_0 + G_1)$$

It is seen that the both sides of (43) vanish after scalar multiplication of this equation by $\underline{n}$. Relaxation character of this equation is caused by the "bulk" relaxation due to the viscoelastic nematic character of evolution equation (40) for the transient elastic strain.

Note that in particular case when $\lambda_1 = \lambda_2 \equiv \lambda$, the evolution equation (43) degenerates into the Ericksen equation, $\overset{0}{\underline{n}} = \lambda \mathbf{b}(\underline{n}) \bullet \underline{\underline{e}}$.

## 5. Example: Symmetric linear nematic viscoelasticity

The above CE's for nematic viscoelasticity are weakly nonlinear. The linear CE's can be introduced only when $\underline{n} = const$. They allow exact solutions in both non-symmetric and symmetric *monodomain* cases, when the N-operator could depend only on time $t$. We demonstrate below only a simple solution for symmetric case, using the N-operator technique.

When $\underline{n} = const$ the closed set of CE's has the form:

$$\dot{\underline{\underline{\varepsilon}}} + \mathbf{s}(\underline{n}) \bullet \underline{\underline{\varepsilon}} = \underline{\underline{e}}, \quad \underline{\underline{\sigma}} = \mathbf{G}(\underline{n}) \bullet \underline{\underline{\varepsilon}}. \tag{$45_{1,2}$}$$

Here $\mathbf{G}(\underline{n}), \mathbf{s}(\underline{n}) \in \check{\mathrm{S}}_3$ have the TI-operator forms, $\mathbf{G}(\underline{n}) = \sum_{k=0}^{2} \hat{G}_k \mathbf{a}_k(\underline{n})$, $\mathbf{s}(\underline{n}) = \sum_{i=0}^{2} s_i \mathbf{a}_i(\underline{n})$, where $\hat{G}_k$ and $s_k$ are the basis scalars of moduli and relaxation frequency denoted in ($33_1$) and ($39_2$). Searching the solution of homogeneous equation in ($45_1$) in the form, $\underline{\underline{\varepsilon}} = \underline{\underline{\psi}} \exp(-\nu t)$, reduces it to the common spectral problem with eigenvalue $\nu$ and "eigentensor" $\underline{\underline{\psi}} \in \breve{A}^s$ as:

$$\mathbf{s}(\underline{n}) \bullet \underline{\underline{\psi}} - \nu \underline{\underline{\psi}} \equiv \tilde{\mathbf{s}}(\nu, \underline{n}) \bullet \underline{\underline{\psi}} = \underline{\underline{0}}, \quad \tilde{\mathbf{s}}(\nu, \underline{n}) \equiv \mathbf{s}(\underline{n}) - \nu \mathbf{a}_0 \quad (\underline{\underline{\psi}} \in \breve{A}^s) \tag{46}$$

The eigenvalues $\nu_k$, being the spectral points of operator $\tilde{\mathbf{s}}(\underline{n})$, are the singular values of inverse TI-operator $\tilde{\mathbf{s}}^{-1}(\underline{n})$. It means that they are the singular points of the basis parameters defined in (A18), where $\hat{s}_k = \hat{r}_k, r_0 = s_0 + \nu$, $r_1 = s_1$, $r_2 = s_2$. Brief inspection of these dependencies results in the formulae for eigenvalues:

$$\nu_1 = s_0 = \frac{G_0}{\eta_0}, \quad \nu_2 = s_0 + s_1 = \frac{G_0 + G_1}{\eta_0 + \eta_1}, \; \nu_3 = s_0 + \frac{2}{3} s_2 = \frac{3/4 G_0 + G_1 + G_2}{3/4 \eta_0 + \eta_1 + \eta_2}. \tag{47}$$

Due to the thermodynamic stability conditions ($11_{1,2}$) all the eigenvalues $\nu_k$ are positive and describe the *relaxation frequencies*, with the respective *relaxation times* $\hat{\theta}_k = 1/\nu_k$.

Instead of determining the eigentensors $\underline{\underline{\psi}}$ of the problem (46), we now search for the "4-eigentensors" $\mathbf{Q}(\nu,\underline{n})$, as nontrivial solutions of the operator problem:

$$\mathbf{s}(\underline{n})\bullet\mathbf{Q}-\nu\mathbf{Q}\equiv\tilde{\mathbf{s}}(\nu,\underline{n})\bullet\mathbf{Q}=\mathbf{0}\,,\quad \mathbf{Q}(\nu,\underline{n})=\sum_{k=0}^{2}q_k(\nu)\mathbf{a}_k(\underline{n})\,. \tag{48}$$

If the "4-eigentensors" $\mathbf{Q}(\nu,\underline{n})$ are found, the eigentensor $\underline{\underline{\psi}}$ is determined as: $\underline{\underline{\psi}}=\mathbf{Q}(\nu,\underline{n})\bullet\underline{\underline{\psi}}_0$, where the tensor $\underline{\underline{\psi}}_0\in\breve{A}^s$ is given.

For each $\nu_k$ defined in (47), the basis parameters $q_k(\nu)$ are now found from (A15), where all $t_k=0$, $r_0=s_0+\nu$, $r_1=s_1$, and $r_2=s_2$,. Easy calculations show that

$$\{q(\nu_1)\}=c_1(1,-1,-3/2),\quad \{q(\nu_2)\}=c_2(0,1,0)\,,\quad \{q(\nu_3)\}=c_3(0,0,1)\,. \tag{49}$$

Therefore the 4-eigentensors $\mathbf{Q}(\nu,\underline{n})$ are:

$$\mathbf{Q}(\nu_k,\underline{n})=c_k\mathbf{Q}_k(\underline{n});\qquad \mathbf{Q}_1(\underline{n})=\mathbf{a}_0-\mathbf{a}_1-3/2\mathbf{a}_2\,,\quad \mathbf{Q}_2(\underline{n})=\mathbf{a}_1\,,\quad \mathbf{Q}_3(\underline{n})=\mathbf{a}_2\,. \tag{50}$$

Using the routine procedure, the solution of evolution equation ($45_1$) is found as:

$$\underline{\underline{\varepsilon}}(t)=\sum_{k=1}^{3}c_k\mathbf{Q}_k(\underline{n})\int_{-\infty}^{t}\underline{\underline{e}}(\tau)\exp[-\nu_k(t-\tau)]d\tau\qquad (c_1=c_2=1,\ \ c_3=3/2), \tag{51}$$

Here the values of $c_k$ have been determined from the equation, $\sum_{k=1}^{3}c_k\mathbf{Q}_k(\underline{n})=\mathbf{a}_0$, obtained. by substituting (51) into ($45_1$), with the use of (50). The stress-strain rate CE's, found from ($45_2$) with the aid of formulae ($33_1$) and (A15), have the TI-operator presentation:

$$\underline{\underline{\sigma}}(t)=\sum_{k=0}^{2}\mathbf{a}_k(n)\bullet\underline{\underline{E}}_k(t),\quad \underline{\underline{E}}_k(t)\equiv\int_{-\infty}^{t}m_k(t-\tau)\underline{\underline{e}}(\tau)d\tau\,. \tag{$52_1$}$$

Here

$$m_0(t)=G_0e^{-\nu_1 t},\ \ m_1(t)=G_1e^{-\nu_2 t}+G_0(e^{-\nu_2 t}-e^{-\nu_1 t}),\ \ m_2(t)=2(G_1+G_2)e^{-\nu_3 t}+\frac{3}{2}G_0(e^{-\nu_3 t}-e^{-\nu_1 t}) \tag{$52_2$}$$

The common tensor expression of ($52_1$) is:

$$\underline{\underline{\sigma}}(t)=\underline{\underline{E}}_0(t)+\underline{n}\underline{n}\cdot\underline{\underline{E}}_1(t)+\underline{\underline{E}}_1(t)\cdot\underline{n}\underline{n}-2\underline{n}\underline{n}[\underline{\underline{E}}_1(t):\underline{n}\underline{n}]+[\underline{\underline{E}}_2(t):\underline{n}\underline{n}](\underline{n}\underline{n}-1/3\underline{\underline{\delta}}). \tag{$52_3$}$$

Here the linear memory functionals $\underline{\underline{E}}_k(t)$ are defined in ($52_{1,2}$). The CE ($52_3$) derived here from nematodynamic CE's ($45_{1,2}$) is of type, proposed by Larson and Mead (1989)

There are two physically evident limits of the memory functionals $\underline{\underline{E}}_k(t)$.

1) When $t << \min \nu_k^{-1} = \min \hat{\theta}_k$, $\underline{\underline{E}}_k(t) \approx m_k(0)\int_{-\infty}^{t} \underline{\underline{e}}(\tau)d\tau = \hat{G}_k \underline{\underline{\varepsilon}}(t)$ where $\hat{G}_k$ are the basis scalars of TI moduli operator $\mathbf{G}(\underline{n})$ defined in ($33_1$). This is the purely elastic limit of ($52_3$) which coincides with ($28_1$);

2) When $t \to \infty$ and the constant limit $\underline{\underline{e}} \equiv \underline{\underline{e}}(\infty)$ exists, $\underline{\underline{E}}_k(\infty) = \underline{\underline{e}}\int_0^{\infty} m_k(t)dt = \hat{\eta}_k \underline{\underline{e}}$, where $\hat{\eta}_k$ are basis scalars of the TI viscosity operator $\boldsymbol{\eta}(\underline{n})$, defined in ($33_2$). In this case the relations ($52_3$) are purely viscous and coincide with ($28_2$), where $\underline{\underline{e}}_p \to \underline{\underline{e}}$.

## Conclusions

The present paper develops a continual constitutive theory for weak viscoelastic nematodynamics of Maxwell type, using the standard local approach of non-equilibrium thermodynamics. Weakly nonlinear character of this theory is based on the assumption of smallness of elastic strains and relative rotations, meaning in fact the smallness of the Deborah numbers, with valid in this case the co-rotational derivatives. Along with specific viscoelastic and nematic kinematical relations, the theory utilized the CE’s resulting from the de Gennes potential for weakly elastic nematic solids, extended for the presence of magnetic field, and the LEP CE’s for viscous nematic liquids, while ignoring the Frank (orientation) elasticity and inertia effects. In spite of many constitutive parameters involved in the non-symmetric theory, a non-commutative group of linear nematic operations, introduced and analyzed in Appendix, allowed us to reveal a general, simple structure of the theory. It is shown that in non-symmetric case, the N-operators of elastic moduli $\mathbf{G}(\underline{n})$ and viscosity $\boldsymbol{\eta}(\underline{n})$ are non-commutative. It was also shown that the evolution equation for director is of viscoelastic type. The stress asymmetry is demonstrated on the example of magnetic field acted on LCP. In the case when stress tensor is symmetric, the theory is simplified. The rheological properties in symmetric case are described by a commutative group of nematic transformations. We also demonstrated that in linear symmetric case the CE’s are reduced to the type proposed by Larson and Mead (1989) .

We think that even in the general multi-parametric form, the present theory could serve as a fundamental basis for more detailed but particular theoretical approaches.

As found in our former publications for nematic elastic solids and viscous liquids [Leonov and Volkov (2004a,b)], the existence of possible soft deformation modes in nematic theories highly reduces the number of constitutive parameters. Such a theory for possible *viscoelastic* nematic soft modes has yet to be developed.

The large volume of this paper does not allow us to discuss the specific features of the theory for simple flows, such as simple shearing and simple elongation. It will be done in a separate paper along with comparing our predictions with the experimental data.

## Appendix: Nematic fourth and third rank tensors and operations

We introduce here the anisotropic third and fourth rank "nematic" tensors and describe basic algebraic operations with them. These tensors occur first in the Section 2 for describing nematic liquids with uniaxial symmetry of anisotropic properties. In order to reveal the complicated mathematical structure of the nematic tensor operations we make several steps to overcome formal obstacles, the most unpleasant being the cumbersomeness of the standard tensor techniques. In the first step we explicitly introduce the "basis" numerical tensors, expose their symmetry properties and demonstrate that the set of basis numerical tensors is irreducible. It means that each nematic fourth rank tensor (or "4-tensor") as those introduced in the main text, under certain conditions is uniquely represented via the basis tensors. Finally, we establish that under the same conditions the set of nematic operators represented by the 4-tensors forms a group with respect to their multiplicative properties. The use of 4-tensors simplifies the derivations of the evolution equations for tensor state variables. Additionally, we introduce in Appendix a third rank numerical tensor, (so called "3-tensor") and basic operations where this tensor participates. The use of 3-tensors simplifies the derivations of the evolution equations for director. All of this allows us to introduce in the main text the operator formulations of CE's with simplified structure and notations. We consider below the structure properties only for Cartesian 3- and 4-tensors.

## *A1. Definitions and properties of symmetry*

The *basis* numerical 4-tensors describing the nematic anisotropy are defined as:

$$\{\mathbf{a}_0\}_{ij\alpha\beta} = a^{(0)}_{ij\alpha\beta} = 1/2(\delta_{i\alpha}\delta_{j\beta} + \delta_{i\beta}\delta_{j\alpha} - 2/3\delta_{ij}\delta_{\alpha\beta}) \quad (\mathrm{A1}_1)$$

$$\{\mathbf{a}_1(\underline{n})\}_{ij\alpha\beta} = a^{(1)}_{ij\alpha\beta} = 1/2(\delta^{\perp}_{i\alpha}n_j n_\beta + \delta^{\perp}_{j\alpha}n_i n_\beta + \delta^{\perp}_{i\beta}n_j n_\alpha + \delta^{\perp}_{j\beta}n_i n_\alpha) \quad (\mathrm{A1}_2)$$

$$\{\mathbf{a}_2(n)\}_{ij\alpha\beta} = a^{(2)}_{ij\alpha\beta} = (n_i n_j - 1/3\delta_{ij})(n_\alpha n_\beta - 1/3\delta_{\alpha\beta}) \quad (\mathrm{A1}_3)$$

$$\{\mathbf{a}_3(\underline{n})\}_{ij\alpha\beta} = a^{(3)}_{ij\alpha\beta} = 1/2(\delta_{i\alpha}n_j n_\beta + \delta_{j\alpha}n_i n_\beta - \delta_{i\beta}n_j n_\alpha - \delta_{j\beta}n_i n_\alpha) \quad (\mathrm{A1}_4)$$

$$\{\mathbf{a}_4(\underline{n})\}_{ij\alpha\beta} = a^{(4)}_{ij\alpha\beta}(\underline{n}) = 1/2(\delta_{i\alpha}n_j n_\beta + \delta_{i\beta}n_j n_\alpha - \delta_{j\alpha}n_i n_\beta - \delta_{j\beta}n_i n_\alpha) = \{\mathbf{a}_3(\underline{n})\}_{\alpha\beta ij} \quad (\mathrm{A1}_5)$$

$$\mathbf{a}_5(\underline{n})_{ij\alpha\beta} = a^{(5)}_{ij\alpha\beta}(\underline{n}) = 1/2(\delta_{i\beta}n_j n_\alpha - \delta_{i\alpha}n_j n_\beta + \delta_{j\alpha}n_i n_\beta - \delta_{j\beta}n_i n_\alpha) \quad (\mathrm{A1}_6)$$

Additionally, we also introduce the numerical 3-tensor,

$$\{\mathbf{b}(\underline{n})\}_{ij\nu} = b_{ij\nu}(\underline{n}) = \delta^{\perp}_{ij}n_\nu \quad , \quad \delta^{\perp}_{ij} = \delta_{ij} - n_i n_j \, . \quad (\mathrm{A1}_7)$$

The 4-tensors defined in ($\mathrm{A1}_{1\text{-}6}$) are traceless with respect of the first and the second pairs of indices, i.e. $\{\mathbf{a}_k(\underline{n})\}_{ii\alpha\beta} = \{\mathbf{a}_k(\underline{n})\}_{ij\alpha\alpha} = 0$ ($k = 0,..,5$).

These tensors have the *symmetry properties* that follow from the symmetry properties of the stress and existence of free energy (dissipation):

$$\{\mathbf{a}_k(\underline{n})\}_{ij\alpha\beta} = \{\mathbf{a}_k(\underline{n})\}_{ji\alpha\beta} = \{\mathbf{a}_k(\underline{n})\}_{ij\beta\alpha} = \{\mathbf{a}_k(\underline{n})\}_{\alpha\beta ij} \qquad (k = 0,1,2) \quad (\mathrm{A2}_1)$$

$$\{\mathbf{a}_3(\underline{n})\}_{ij\alpha\beta} = \{\mathbf{a}_3(\underline{n})\}_{ji\alpha\beta} = -\{\mathbf{a}_3(\underline{n})\}_{ij\beta\alpha} \quad (\mathrm{A2}_2)$$

$$\{\mathbf{a}_4(\underline{n})\}_{ij\alpha\beta} = -\{\mathbf{a}_3(\underline{n})\}_{ji\alpha\beta} = \{\mathbf{a}_3(\underline{n})\}_{ij\beta\alpha} \quad (\mathrm{A2}_3)$$

$$\{\mathbf{a}_5(\underline{n})\}_{ij\alpha\beta} = -\{\mathbf{a}_5(\underline{n})\}_{ji\alpha\beta} = -\{\mathbf{a}_5(\underline{n})\}_{ij\beta\alpha} = \{\mathbf{a}_5(\underline{n})\}_{\alpha\beta ij} \, . \quad (\mathrm{A2}_4)$$

Formulae ($\mathrm{A2}_{1\text{-}4}$) show that the first three 4-tensors $\{\mathbf{a}_k(\underline{n})\}_{ij\alpha\beta}$ (k = 0,1,2) are symmetric under transposition of the first and second indices, the third and fourth indices, as well as the first and second pairs of indices; the tensor $\{\mathbf{a}_3(\underline{n})\}_{ij\alpha\beta}$ is symmetric under transposition of the first and second indices and skew symmetric under transposition of the third and fourth indices; the tensor $\{\mathbf{a}_4(\underline{n})\}_{ij\alpha\beta}$ is asymmetric under transposition of the first and second indices and symmetric with transposition of the third and fourth indices; and the tensor $\{\mathbf{a}_5(\underline{n})\}_{ij\alpha\beta}$ is asymmetric with transposition of the first and second, as well as of the third and fourth indices, and symmetric with transposition of the first and second pairs of indices. The above properties of basis 4-tensors $\mathbf{a}_k(\underline{n})$ $(k = 0,..,5)$ show that they represent irreducible set of traceless 4-th rank tensors. Therefore we call them *basis* tensors.

Evidently, the 3-tensor $\mathbf{b}(\underline{n})$ defined in (A1$_7$) is symmetric under transposition of the first two indices.

### *A2. Multiplicative properties of numerical tensors*

The matrix products of tensors of different ranks are defined as:

$$\mathbf{a}_s(\underline{n})\bullet\mathbf{a}_r(\underline{n}) \Rightarrow a^{(s)}_{ij\alpha\beta}a^{(r)}_{\beta\alpha\nu\gamma},\ \ \mathbf{a}_s(\underline{n})\bullet\mathbf{b}(\underline{n}) \Rightarrow a^{(s)}_{ij\alpha\beta}b_{\beta\alpha\nu}$$
$$\mathbf{a}_s(\underline{n})\bullet\underline{\underline{x}} \Rightarrow a^{(s)}_{ij\alpha\beta}x_{\beta\alpha},\ \ \mathbf{b}(\underline{n})\bullet\underline{\underline{x}} \Rightarrow b_{i\alpha\beta}x_{\beta\alpha},\ \text{etc.} \qquad \text{(A3)}$$

The products $\mathbf{a}_s(\underline{n})\bullet\mathbf{a}_r(\underline{n})$, established directly are presented in Table 1.

Table 1. Products of basis tensors $\mathbf{a}_i\bullet\mathbf{a}_j$; the first index *i* showing the rows, the second *j* showing the columns

| ***j*** / ***i*** | 0 | 1 | 2 | 3 | 4 | 5 |
|---|---|---|---|---|---|---|
| 0 | $\mathbf{a}_0$ | $\mathbf{a}_1$ | $\mathbf{a}_2$ | $\mathbf{a}_3$ | $\mathbf{0}$ | $\mathbf{0}$ |
| 1 | $\mathbf{a}_1$ | $\mathbf{a}_1$ | $\mathbf{0}$ | $\mathbf{a}_3$ | $\mathbf{0}$ | $\mathbf{0}$ |
| 2 | $\mathbf{a}_2$ | $\mathbf{0}$ | $(2/3)\,\mathbf{a}_2$ | $\mathbf{0}$ | $\mathbf{0}$ | $\mathbf{0}$ |
| 3 | $\mathbf{0}$ | $\mathbf{0}$ | $\mathbf{0}$ | $\mathbf{0}$ | $-\mathbf{a}_1$ | $\mathbf{a}_3$ |
| 4 | $\mathbf{a}_4$ | $\mathbf{a}_4$ | $\mathbf{0}$ | $-\mathbf{a}_5$ | $\mathbf{0}$ | $\mathbf{0}$ |
| 5 | $\mathbf{0}$ | $\mathbf{0}$ | $\mathbf{0}$ | $\mathbf{0}$ | $\mathbf{a}_4$ | $\mathbf{a}_5$ |

It is seen that except *i,j* = 0,1,2, the operation of multiplication of basis tensors is non-commutative, for example, $\mathbf{a}_0\bullet\mathbf{a}_3 = \mathbf{a}_3 \neq \mathbf{a}_3\bullet\mathbf{a}_0 = \mathbf{0}$.

Additionally, there are the following operations involving 3-tensor $\mathbf{b}(\underline{n})$:

$$\mathbf{b}\bullet\mathbf{a}_0 = \mathbf{b}\bullet\mathbf{a}_1 = \mathbf{b}^s,\ \mathbf{b}\bullet\mathbf{a}_2 = \mathbf{0},\ \mathbf{b}\bullet\mathbf{a}_3 = \mathbf{b}^a,\ \mathbf{b}\bullet\mathbf{a}_4 = -\mathbf{b}^s,\ \mathbf{b}\bullet\mathbf{a}_5 = \mathbf{b}^a\ . \qquad \text{(A4)}$$

If $\underline{\underline{x}}^{\mathrm{s}}$ and $\underline{\underline{x}}^{\mathrm{a}}$ are respectively symmetric and asymmetric tensors then

$$\mathbf{b}^s\bullet\underline{\underline{x}}^s = \mathbf{b}\bullet\underline{\underline{x}}^s,\ \ \mathbf{b}^a\bullet\underline{\underline{x}}^a = \mathbf{b}\bullet\underline{\underline{x}}^a = -\underline{\underline{x}}^a\cdot\underline{n},\ \ b^s_{ijk} = 1/2(b_{ijk}+b_{ikj}),\ \ b^a_{ijk} = 1/2(b_{ijk}-b_{ikj}) \qquad \text{(A5)}$$

### *A3. Non-symmetric linear nematic operators and their algebraic properties*

Consider the linear space $\breve{X}$ of traceless second rank tensors $\underline{\underline{x}} \in \breve{X}$: $tr\underline{\underline{x}} = 0$. Every tensor $\underline{\underline{x}} \in \breve{X}$ is uniquely represented as the sum, $\underline{\underline{x}} = \underline{\underline{x}}^s + \underline{\underline{x}}^a$, where $\underline{\underline{x}}^s$ and $\underline{\underline{x}}^a$ are symmetric and asymmetric tensors in $\breve{X}$, respectively. We introduce on $\breve{X}$ the linear nematic operation or simply N-operator, $\mathbf{N}_r(\underline{n})$, represented by the 4-tensor as:

$$\underline{\underline{y}} = \mathbf{N}_r(\underline{n}) \bullet \underline{\underline{x}}, \qquad \mathbf{N}_r(\underline{n}) \equiv \sum_{k=0}^{5} r_k \mathbf{a}_k(\underline{n}) \qquad (\mathrm{A6}_1)$$

Here $\mathbf{a}_k(\underline{n})$ are the *basis tensors* defined in (A1), and $r_k$ are the real–valued *basis scalars* characterized operation, which are described for convenience as the ordered set, $\{r\} = (r_0, r_1, r_2, r_3, r_4, r_5)$. Using the equality $\underline{\underline{x}} = \underline{\underline{x}}^s + \underline{\underline{x}}^a$ and symmetry properties (A2), the first relation in (A6$_1$) is rewritten as:

$$\underline{\underline{y}}^s = \sum_{k=0}^{2} r_k \mathbf{a}_k(\underline{n}) \bullet \underline{\underline{x}}^s + r_3 \mathbf{a}_3(\underline{n}) \bullet \underline{\underline{x}}^a, \quad \underline{\underline{y}}^a = r_4 \mathbf{a}_4(\underline{n}) \bullet \underline{\underline{x}}^s + r_5 \mathbf{a}_5(\underline{n}) \bullet \underline{\underline{x}}^a \ . \qquad (\mathrm{A6}_2)$$

A particular version of $\mathbf{N}_r$(n) when there is the Onsager-type symmetry, $r_4 = -r_3$ , is called Onsager nematic operation, or simply ON-operator denoted as $\mathbf{N}_r^o(\underline{n})$. If there are two consequent N-operations, $\underline{\underline{y}} = \mathbf{N}_r(\underline{n}) \bullet \underline{\underline{x}}$ and $\underline{\underline{z}} = \mathbf{N}_q(\underline{n}) \bullet \underline{\underline{y}}$, the resulting N-operation is presented as: $\underline{\underline{z}} = \mathbf{N}_t(\underline{n}) \bullet \underline{\underline{x}}$. Here

$$\mathbf{N}_t(\underline{n}) \equiv \mathbf{N}_q(\underline{n}) \bullet \mathbf{N}_r(\underline{n}) = \sum_{k,m=0}^{5} q_k r_m \mathbf{a}_k(\underline{n}) \bullet \mathbf{a}_m(\underline{n}) = \sum_{k=0}^{6} t_k \mathbf{a}_k(\underline{n}). \qquad (\mathrm{A7})$$

Using the multiplicative Table 1, the basic scalars $t_\mathrm{k}$ of resulting operation are defined from the characteristic equation:

$$\begin{aligned} &t_0 = r_0 q_0, \ t_1 = r_0 q_1 + r_1 q_0 + r_1 q_1 - r_4 q_3, \ t_2 = r_0 q_2 + r_2 q_0 + 2/3 r_2 q_2 \\ &t_3 = (q_0 + q_1) r_3 + q_3 r_5, \ t_4 = (r_0 + r_1) q_4 + r_4 q_5, \ t_5 = -r_3 q_4 + r_5 q_5 \end{aligned} \qquad (\mathrm{A8})$$

Even in the Onsager case, when $q_4 = -q_3$, and $r_4 = -r_3$, generally $t_4 \neq -t_3$, i.e. $\mathbf{N}_t(\underline{n}) \neq \mathbf{N}_t^o(\underline{n})$ meaning that the product of two ON-operators is not a ON-operator. It should be mentioned that the set of operators $\mathbf{N}_r(\underline{n})$ constitutes the *associative ring* $\mathsf{N}_6$ relative to both addition and multiplication operations of elements $\mathbf{N}_r(\underline{n})$ of this set.

A N-operator $\mathbf{N}_r(\underline{n})$ is called *positively definite* if $\forall\ \underline{\underline{x}} \in \breve{X}$ the scalar $P = \underline{\underline{x}} \bullet \mathbf{N}_r(\underline{n}) \bullet \underline{\underline{x}} > 0$. We denote the set of positively definite N-operators as $\check{\mathrm{N}}_6 \subset \mathrm{N}_6$. The analysis of this positive definiteness easily performed in a special Cartesian coordinate system where $\underline{n} = \{1,0,0\}$, is quite similar to that made by Leonov and Volkov (2004a,b), and results in the inequalities for basis scalars $r_k$:

$$\{r\} \in R_+ : \quad r_0 > 0,\ r_0 + 2/3 r_2 > 0,\ r_5(r_0 + r_1) + r_4 r_3 > 0;\ r_0 + r_1 > 0\ (r_5 > 0). \qquad \text{(A9)}$$

Here $R_+$ is the 6-dimensional set of ordered parameters $\{r\}$ satisfying (A9). To avoid degeneration of algebraic relations we hereafter assume that $r_k \neq 0\ (k = 0,1,2,3,4,5)$. The particular case of positively defined ON operators is specified when in (A9), $r_4 = -r_3$ and $R_+^o \equiv R_+\big|_{r_4 = -r_3}$, with particular inequalities in (A9) coinciding with the stability conditions discussed in the main text.

There is a special *unit* N-operator $\mathbf{I}(\underline{n})$. Its definition and basic properties due to the Table A1 are:

$$\mathbf{I}(\underline{n}) = \mathbf{a}_0(\underline{n}) + \mathbf{a}_5(\underline{n}), \quad \mathbf{N}_q(\underline{n}) \bullet \mathbf{I}(\underline{n}) = \mathbf{I}(\underline{n}) \bullet \mathbf{N}_q(\underline{n}) = \mathbf{N}_q(\underline{n}) \qquad \text{(A10)}$$

Consider a N-operator $\mathbf{N}_r(\underline{n})$, which describes the transformation $\underline{\underline{y}} = \mathbf{N}_r(\underline{n}) \bullet \underline{\underline{x}}$. The inverse N-operator, $\mathbf{N}_r^{-1}(\underline{n}) \equiv \mathbf{N}_{\hat{r}}(\underline{n})$, if exists, should satisfy the condition,

$$\mathbf{N}_{\hat{r}}(\underline{n}) \bullet \mathbf{N}_r(\underline{n}) = \mathbf{N}_r(\underline{n}) \bullet \mathbf{N}_{\hat{r}}(\underline{n}) = \mathbf{I}(\underline{n}).$$

Satisfying this condition with the aid of (A8) yields:

$$t_0 = 1,\ t_1 = t_2 = t_3 = t_4 = 0,\ t_5 = 1. \qquad \text{(A11)}$$

The basis scalars $\hat{r}_k$ of inverse N-operator $\mathbf{N}_{\hat{r}}(\underline{n})$ are found using (A11) and (A.9) as:

$$\hat{r}_0 = \frac{1}{r_0}, \quad \hat{r}_1 = \frac{-(r_3 r_4 + r_1 r_5)/r_0}{r_5(r_0 + r_1) + r_3 r_4}, \quad \hat{r}_2 = \frac{-r_2/r_0}{r_0 + 2/3 r_2}$$
$$\hat{r}_3 = \frac{-r_3}{r_5(r_0 + r_1) + r_3 r_4}, \quad \hat{r}_4 = \frac{-r_4}{r_5(r_0 + r_1) + r_3 r_4}, \quad \hat{r}_5 = \frac{r_0 + r_1}{r_5(r_0 + r_1) + r_3 r_4} \qquad \text{(A12)}$$

In case of ON-operators, when $r_4 = -r_3$, relations (A.12) yield: $\hat{r}_4 = -\hat{r}_3$, i.e. $\mathbf{N}_r^{o-1}(\underline{n}) = \mathbf{N}_{\hat{r}}^o(\underline{n})$, meaning that inverse ON-operator is still ON-operator. Applying inequalities (A9) to formulae (A12) one can see that each N-operator from the set $\check{\mathrm{N}}_6$ always has its inverse. One can also prove that the product of two positively definite nematic operators is positively definite. All these properties mean that the set $\check{N}_6$ constitutes a *six-parametric non-commutative group* of N-operators with the *characteristic group equation* (A9). The elements of this group $\mathbf{N}_r(\underline{n}) \in \check{N}_6$ have the *4-tensor presentation* shown in the second equation ($A6_1$). The group $\check{N}_6$ is non-trivial because it consists of six independent linear nematic operations, described by the basis 4-tensors shown in (A1). Note that the class $\check{N}_6^o$ of positively definite ON operators does not constitute a subgroup in $\check{N}_6$, because their products do not belong to $\check{N}_6^o$.

The properties of $\check{N}_6$ group highly simplify transformations of CE's in the main text, reducing them only to the analysis of characteristic group equation (A9).

### *A4. Symmetric case: transversal-isotropic (TI) operations*

Consider herewith only the sub-space $\breve{X}^s \subset \breve{X}$ of traceless second rank symmetric tensors $\underline{\underline{x}}^s \in \breve{X}^s$ : $tr\underline{\underline{x}}^s = 0$. A linear symmetric TI-operator $\mathbf{S}_r(\underline{n})$ on $\breve{X}^s$ presented by its 4-tensor is defined as:

$$\underline{\underline{y}}^s = \mathbf{S}_r(\underline{n}) \bullet \underline{\underline{x}}^s = \sum_{k=0}^{2} r_k \mathbf{a}_k(\underline{n}) \bullet \underline{\underline{x}}^s, \quad \text{or} \quad \mathbf{S}_r(\underline{n}) = \sum_{k=0}^{2} r_k \mathbf{a}_k(\underline{n}) \tag{A13}$$

Here $\mathbf{a}_k(\underline{n})$ are the basis tensors defined in (A1), and $r_k$ are the real-valued basic scalar ordered parameters $\{r\}$, characterizing operation. We consider further only the non-degenerating situation when all $r_k \neq 0$ ($k$ = 0,1,2). As follows from Table A1, the products of TI-operators are commutative and are presented as linear combinations of $\mathbf{a}_k(\underline{n})$ ($k$ = 0,1,2). In particular,

$$\mathbf{S}_r(\underline{n}) \bullet \mathbf{S}_q(\underline{n}) = \mathbf{S}_q(\underline{n}) \bullet \mathbf{S}_r(\underline{n}) = \mathbf{S}_t(\underline{n}), \tag{A14}$$

where the basis parameters $t_k$ are found from the characteristic equation:

$$t_0 = r_0 q_0, \;\; t_1 = r_0 q_1 + r_1 q_0 + r_1 q_1, \;\; t_2 = r_0 q_2 + r_2 q_0 + 2/3 r_2 q_2. \tag{A15}$$

The unit operation here is $\mathbf{I} = \mathbf{a}_0$.

A TI-operator presented by its 4-tensor $\mathbf{S}_r(\underline{n}) \in \check{S}_3$ is *called positively definite* if $\forall \underline{\underline{x}}^s \in \breve{X}^s$ the quadratic form $P^s = \underline{\underline{x}}^s \bullet \mathbf{S}_r(\underline{n}) \bullet \underline{\underline{x}}^s > 0$. It happens if and only if

$$r_0 > 0, \ r_0 + r_1 > 0, \ r_0 + 2/3 r_2 > 0. \qquad \text{(A16)}$$

Direct calculations show that a product of two positively definite TI-operators is positively definite.

The scalar basis parameters $\hat{r}_k$ for inverse TI-operator $\mathbf{S}_r^{-1}(\underline{n}) \equiv \mathbf{S}_{\hat{r}}(\underline{n})$ (if exists) are found from the relation, $\mathbf{S}_r^{-1}(\underline{n}) \bullet \mathbf{S}_r(\underline{n}) = \mathbf{S}_{\hat{r}}(\underline{n}) \bullet \mathbf{S}_r(\underline{n}) = \mathbf{a}_0(\underline{n})$, which reduces (A.15) to:

$$t_0 = r_0 \hat{r}_0 = 1, \ t_1 = r_0 \hat{r}_1 + r_1 \hat{r}_0 + r_1 \hat{r}_1 = 0, \ t_2 = r_0 \hat{r}_2 + r_2 \hat{r}_0 + 2/3 r_2 \hat{r}_2 = 0. \qquad \text{(A17)}$$

Thus the basis scalar parameters of inverse TI operator (if exist) are given by:

$$\hat{r}_0 = \frac{1}{r_0}, \ \hat{r}_1 = -\frac{r_1 / r_0}{r_0 + r_1}, \ \hat{r}_2 = -\frac{r_2 / r_0}{r_0 + 2/3 r_2}. \qquad \text{(A18)}$$

As follows from (A16), the inverse TI-operator $\mathbf{S}_r^{-1}(\underline{n}) \equiv \mathbf{S}_{\hat{r}}(\underline{n})$ always exists for any positively definite TI-operator $\mathbf{S}_r(\underline{n}) \in \check{S}_3$. It means that $\check{S}_3$ is a commutative three parametric TI group. Evidently $\check{S}_3 \subset \check{N}_6$.